\documentclass[aps,prd,preprintnumbers,showpacs,superscriptaddress,nofootinbib,amsmath,amssymb,floats,floatfix,showkeys,notitlepage,longbibliography]{revtex4-1}
\usepackage{comment}
\usepackage{graphicx}
\usepackage{subfigure}
\usepackage{palatino}
\usepackage[commandnameprefix=always]{changes}
\usepackage{hyperref}
\hypersetup{colorlinks=true,linkcolor=red,urlcolor=red,citecolor=red}
\usepackage[toc,page]{appendix}
\usepackage[normalem]{ulem}

\usepackage{lipsum}
\usepackage{graphicx}
\usepackage{subfigure}
\usepackage{palatino}
\usepackage{float}
\usepackage{sans}
\usepackage{adjustbox}
\usepackage{latexsym}
\usepackage{amsmath}
\usepackage{amssymb}
\usepackage{amsfonts}
\usepackage{dcolumn}
\usepackage{bm}
\usepackage{tikz}
\usepackage{bigints}
\usepackage{array,tabularx,multirow,booktabs}
\usepackage[tracking=true]{microtype}
\SetTracking{}{500}
\SetTracking{encoding={*}, shape=sc}{40}
\allowdisplaybreaks
\usepackage{adjustbox}
\usepackage{latexsym}
\usepackage{amsmath}
\usepackage{amssymb}
\usepackage{amsfonts}
\usepackage{dcolumn}
\usepackage{bm}
\usepackage{tikz}
\usepackage{bigints}
\usepackage{array,tabularx,multirow,booktabs}
\usepackage[tracking=true]{microtype}
\usepackage{color}
\allowdisplaybreaks

\def\jnl@style{\it}
\def\aaref@jnl#1{{\jnl@style#1}}

\def\aaref@jnl#1{{\jnl@style#1}}

\def\aj{\aaref@jnl{AJ}}                   
\def\apj{\aaref@jnl{ApJ}}                 
\def\apjl{\aaref@jnl{ApJ}}                
\def\apjs{\aaref@jnl{ApJS}}               
\def\apss{\aaref@jnl{Ap\&SS}}             
\def\aap{\aaref@jnl{A\&A}}                
\def\aapr{\aaref@jnl{A\&A~Rev.}}          
\def\aaps{\aaref@jnl{A\&AS}}              
\def\mnras{\aaref@jnl{Mon.~Not.~Roy.~Astron.~Soc.}}             
\def\prd{\aaref@jnl{Phys.~Rev.~D}}        
\def\prc{\aaref@jnl{Phys.~Rev.~C}}  
\def\prl{\aaref@jnl{Phys.~Rev.~Lett.}}    
\def\qjras{\aaref@jnl{QJRAS}}             
\def\skytel{\aaref@jnl{S\&T}}             
\def\ssr{\aaref@jnl{Space~Sci.~Rev.}}     
\def\zap{\aaref@jnl{ZAp}}                 
\def\nat{\aaref@jnl{Nature}}              
\def\aplett{\aaref@jnl{Astrophys.~Lett.}} 
\def\apspr{\aaref@jnl{Astrophys.~Space~Phys.~Res.}} 
\def\physrep{\aaref@jnl{Phys.~Rep.}}      
\def\physscr{\aaref@jnl{Phys.~Scr}}       
\def\commat{\aaref@jnl{Comm.~Math.~Phys.}}              
\def\science{\aaref@jnl{Science}}               
\def\cqg{\aaref@jnl{Classical Quant.~Grav.}}            
\def\jpcs{\aaref@jnl{JPCS}}                                     
\def\ijmpd{\aaref@jnl{Int.~J.~Mod.~Phys.~D}}                    
\def\grg{\aaref@jnl{Gen.~Relat.~Gravit.}}               
\def\rpp{\aaref@jnl{Rep.~Prog.~Phys.}}          
\def\npa{\aaref@jnl{Nucl.~Phys.~A}}        
\def\lrr{\aaref@jnl{Living Rev.~Rel.}}                   
\def\jcap{\aaref@jnl{J.~Cosmology Astropart.~Phys.}}    
\def\rmp{\aaref@jnl{Rev.~Mod.~Phys.}}   
\def\epjc{\aaref@jnl{Eur.~Phys.~J.~C}}

\allowdisplaybreaks[1]

\begin{document}

\title{Thermodynamic topology of Black Holes in $F(R)$-Euler-Heisenberg gravity's Rainbow}

\author{Yassine Sekhmani}
\email[Email: ]{sekhmaniyassine@gmail.com}
\affiliation{Center for Theoretical Physics, Khazar University, 41 Mehseti Street, Baku, AZ1096, Azerbaijan.}
\affiliation{Centre for Research Impact \& Outcome, Chitkara University Institute of Engineering and Technology, Chitkara University, Rajpura, 140401, Punjab, India}
\author{Saeed Noori Gashti}
\email[Email: ]{saeed.noorigashti@stu.umz.ac.ir}
\affiliation{School of Physics, Damghan University, P. O. Box 3671641167, Damghan, Iran}
\author{Mohammad Ali S. Afshar}
\email[Email: ]{m.a.s.afshar@gmail.com}
\affiliation{Department of Physics, Faculty of Basic Sciences, University of Mazandaran\\
P. O. Box 47416-95447, Babolsar, Iran}
\affiliation{Canadian Quantum Research Center, 204-3002 32 Ave Vernon, BC V1T 2L7, Canada}
\author{Mohammad Reza Alipour}
\email[Email: ]{mr.alipour@stu.umz.ac.ir}
\affiliation{School of Physics, Damghan University, P. O. Box 3671641167, Damghan, Iran}
\author{Jafar Sadeghi}\email[Email: ]{pouriya@ipm.ir}
\affiliation{Department of Physics, Faculty of Basic Sciences, University of Mazandaran\\
P. O. Box 47416-95447, Babolsar, Iran}
\affiliation{Canadian Quantum Research Center, 204-3002 32 Ave Vernon, BC V1T 2L7, Canada}
\author{Behnam Pourhassan}
\email{b.pourhassan@du.ac.ir}
\affiliation{School of Physics, Damghan University, P. O. Box 3671641167, Damghan, Iran}
\affiliation{Center for Theoretical Physics, Khazar University, 41 Mehseti Street, Baku, AZ1096, Azerbaijan}
\affiliation{Centre for Research Impact \& Outcome, Chitkara University Institute of Engineering and Technology, Chitkara University, Rajpura, 140401, Punjab, India}
\author{ Javlon Rayimbaev} \email[Email: ]{javlon@astrin.uz} \affiliation{Institute of Fundamental and Applied Research, National Research University TIIAME,\\ Kori Niyoziy 39, Tashkent 100000, Uzbekistan} \affiliation{University of Tashkent for Applied Sciences, Str. Gavhar 1, Tashkent 100149, Uzbekistan} \affiliation {Shahrisabz State Pedagogical Institute, Shahrisabz Str. 10, Shahrisabz 181301, Uzbekistan} \affiliation{Tashkent State Technical University, Tashkent 100095, Uzbekistan}.

\begin{abstract}
The topology of black hole thermodynamics is a fascinating area of study that explores the connections between thermodynamic properties and topological features of black holes. It often involves analyzing critical points in the phase diagrams of black holes and assigning topological charges to these points. One significant approach is based on Duan's topological current $\phi$-mapping theory, which introduces the concept of topological charges to critical points in black hole thermodynamics. We successfully derive the field equations for $F(R)$-Euler-Heisenberg theory, providing a framework for studying the interplay between modified gravity and non-linear electromagnetic effects. We obtain an analytical solution for a static, spherically symmetric,  energy-dependent black hole with constant scalar curvature. Also, our analysis of black holes in F(R)-Euler-Heisenberg gravity’s Rainbow reveals significant insights into their topological properties. We identified the total topological charges by examining the normalized field lines along various free parameters. Our findings indicate that the parameters $( R_0 )$ and $( f_{\epsilon} = g_{\epsilon} )$ influence the topological charges.  These results are comprehensively summarized in Table I. In examining the photon sphere within this model, the sign of the parameter \( R_0 \) plays a crucial role in determining whether the model adopts a dS or AdS configuration. An interesting characteristic of this model is that, in its AdS form, it avoids the formation of naked singularity regions, which sets it apart from many other models. Typically, varying parameter values in other models can result in the division of space into regions of black holes and naked singularities. However, this model consistently retains its black hole behavior by featuring an unstable photon sphere, regardless of parameter values within the acceptable range. In its dS form, the behavior of the model’s photon sphere remains consistent with other dS models and does not exhibit unique differences.
\end{abstract}

\date{\today}



\maketitle
\tableofcontents
\section{Introduction}
The greatest cosmological challenges this century are early-time inflation and late-time acceleration. In precise terms, what does inflation refer to, and what does dark energy consist of? Why and how did the early universe and the late universe expand upwards so very fast? What might explain why the evolution of the universe is similar for small and large curvatures? To tackle these matters, Refs. \cite{DeFelice:2010aj,Faraoni:2010pgm,Nojiri:2010wj,Capozziello:2011et,Bamba:2012cp} outline and investigate Einstein's modified theories of gravity as an alternative framework. Among the various modified theories of gravity, the $F(R)$ theory has aroused a great deal of interest because it offers an explanation for a broad series of observations. For instance, the $F(R)$ theory of gravity provides a plausible hypothesis for the existence of dark matte \cite{Abdalla:2020ypg,Nojiri:2006ri} and the universe's rapid expansion \cite{Faraoni:2010pgm,Nojiri:2010wj,Capozziello:2011et,Bamba:2012cp,Joyce:2014kja,Nojiri:2017ncd}. Even more, the $F(R)$ theory can be harnessed to forecast cosmic acceleration or early universe inflation \cite{Capozziello:2002rd,Aditya:2018cmn,Hu:2007nk}, not to mention massive compact objects \cite{Cooney:2009rr,Cheoun:2013tsa,Astashenok:2020cfv,Astashenok:2021peo,Sarmah:2021ule,Kalita:2021zjg}. Another achievement in the context of $F(R)$ gravity is the complete description of all the evolutionary epochs of the universe while maintaining consistency with Newtonian and post-Newtonian approximations \cite{Capozziello:2007ms}. Within the theory of $F(R)$ gravity, $F(R)$ turns out to be an arbitrary function of the scalar curvature $R$.

Since the theory of $F(R)$ gravity may describe a number of cosmological and astrophysical phenomena, it is vital to examine and highlight the coupling of the Euler-Heisenberg (EH) term as a matter source with $F(R)$ gravity’s rainbow by adopting the $F(R)$-EH Lagrangian. Heisenberg and Euler established the non-linear electrodynamics Lagrangian making use of Dirac's electron-positron theory \cite{Heisenberg:1936nmg}. Schwinger remodelled this nonperturbative one-loop Lagrangian in the context of quantum electrodynamics (QED) \cite{Schwinger:1951nm} and found that it could identify the phenomenon of vacuum polarisation successfully. Due to the one-loop nonperturbative QED, the EH Lagrangian has drawn ever-increasing scrutiny to the subject of generalised black hole solutions.  A wealth of intriguing studies on black holes in the context of EH nonlinear electrodynamics are described in \cite{Bronnikov:1979ex,Bronnikov:2000vy,Yajima:2000kw,Corda:2009xd,Ruffini:2013hia,Hendi:2014xia,Maceda:2018zim,Olvera:2019unw}.

In the search for the unification of general relativity and quantum mechanics, a wide range of challenges have arisen. Nowadays, in spite of a few theories such as string theory \cite{Amati:1988tn}, loop quantum gravity \cite{Amelino-Camelia:1996bln}, and space-time foam models \cite{Amelino-Camelia:1997ieq} that have attempted to tackle this conundrum, the question remains devoid of any phenomenological process. A key reason for unifying gravity with quantum theory is the violation of Lorentz invariance, which we might regard as an indispensable criterion for the elaboration of a quantum theory of gravity. Another distinctive factor among unification theories is that they forecast a maximum energy value of the order at the Planck scale that a particle could attain; i.e., if a particle is qualified by the Standard Model of particle physics, then it should conform to the upper limit of the Planck energy, $5.10^{19}\textit{e}V$, which is experimentally backed up by Abraham et al \cite{PierreAuger:2010gfm}. A further alternative to Lorentz symmetry violation involves, in the ultraviolet limit, modifying the standard energy-moment dispersion ratio. This, in fact, has a long history in theories as diverse as discreteness space-time \cite{tHooft:1996ziz}, Horava-Lifshitz \cite{Horava:2009uw,Horava:2009if}, doubly special relativity \cite{Magueijo:2002am,Magueijo:2001cr}, and loop quantum gravity \cite{Amelino-Camelia:1996bln,Amelino-Camelia:2008aez}. In the framework of doubly special relativity, the dispersion relation works out as $E^2f_{\epsilon}^2-p^2g_{\epsilon}^2=m^2$ where $f_{\epsilon}$ and $g_{\epsilon}$ are functions of $\epsilon= E/E_P$, $E$ standing for the energy of the particle used to analyse space-time and $E_P$ for Planck's energy. The functions $f_{\epsilon}$ and $g_{\epsilon}$ are commonly thought of as rainbow functions and are designed for phenomenological purposes \cite{Dehghani:2018svw}. On the other hand, the standard energy dispersion relation gets recurved in the limit $ \lim\limits_{\epsilon\to 0} f_{\epsilon} = \lim\limits_{\epsilon\to 0} g_{\epsilon} = 1$, otherwise known as the infrared limit. As a matter of fact, Linear Lorentz transformations break the invariance of the theory \cite{Garattini:2014rwa}; nevertheless, non-linear Lorentz transformations are those that keep the theory invariant. Moreover, as well as the speed of light being a constant, the Planck energy is also a constant, and it remains impossible for a particle to rise beyond this limit.

A generalisation of doubly special relativity to what is known as rainbow gravity has been suggested by Magueijo and Smolin \cite{Magueijo:2002xx}, where spacetime is mapped to a family of specified parameters in the metric that is parametrised by $\epsilon$. Accordingly, the spacetime geometry is contingent on the energy of the particle testing it, causing a rainbow of the metric. Meanwhile, a number of studies have incorporated Magueijo and Smolin's theory into the understanding of black holes \cite{a1,a2,a3,a4,a5,a6,a7,a8,a9,a10,a11,a12,a13,a14,a15} and into modifying theories of gravity \cite{Garattini:2012ec,Hendi:2016oxk}. A worthy concrete illustration investigation is concerned with the work of M. Momennia et al. \cite{Hendi:2015pwk} exploiting a non-linear electrodynamics (NED) field in the context of the rainbow gravity, resulting in the validity of the first law of thermodynamics in the presence of rainbow functions. Therefore, the rainbow gravity framework has provided the avenue for the consideration of a large number of investigations, such as the derivation considering the dilatonic gravity of an exact black hole solution minimally coupled to the Born-Infeld term with an energy-dependent Liouville-type potential \cite{Hendi:2017ptl}. Similarly, incorporating the quadratic term of the Gauss-Bonnet gravity in the framework of rainbow gravity leads to modifying the appropriate constraints to work out non-singular universes \cite{Hendi:2016tiy}. It is worth noting that the study of thermodynamic properties is considered the main and impacting process regarding rainbow gravity. Thus, since the energy scale depends on the spacetime metric, entropy and temperature also have the same dependence \cite{Dehghani:2018qvn}.
To gain a topological perspective in thermodynamics, one effective approach is to use Duans topological current $\phi$ mapping theory. Wei et al. introduced two distinct methods to study topological thermodynamics based on temperature and the generalized free energy function. The first method involves analyzing the temperature function by eliminating pressure and utilizing the auxiliary and topological parameter $1/ \sin \theta$. The potential is then constructed based on these assumptions. For further study, see \cite{sa1,sa2,sa3,sa4}. The second method assumes that black holes can be considered defects in the thermodynamic parameter space. Their solutions are investigated using the generalized off-shell free energy. In this context, the stability and instability of the obtained black hole solutions are determined by positive and negative winding numbers, respectively. Additionally, the properties of a field configuration can be deduced from the zero points of the field in space \cite{sa5,sa6,sa7,sa8}.
Wei, Liu, and B. Mann asserted that for different branches of a black hole at an arbitrary temperature, the topological charge number, the sum of the winding numbers, is a universal number independent of the black hole parameters. In the range of small and large black holes, the topological charge number depends solely on
the thermodynamic asymptotic behavior of the black hole temperature. Consequently, black holes can be categorized into three different groups based on the number of topological charges \cite{sa5}. Following the introduction of this second method, various black holes were studied to investigate topological charges. C. H. Liu and J. Wang demonstrated that the topological number remains constant, and the charged GB black holes in the AdS space fall into the same category as the RN-AdS, sharing the same topological numbers. However, their results indicated that charge dependence is not the only effective factor \cite{sa9}. Y. Du and Xi. Zhang examined the rotating charged BTZ black hole model and found that there are only two topological classes for BTZ space-time. \cite{sa10}. Using the same method, Wu investigated the structure of charged Lorentzian Taub-NUT space-times and neutral Lorentzian NUT-charged space-times in 4 dimensions in two separate articles. \cite{sa11}. In the second work, Wu found that the presence of the NUT charge as a parameter did not affect the structures in  symptotically flat space-time but had an effect on asymptotically local AdS space-time \cite{sa12}. For further study, you can see \cite{sa13,sa14,sa15,sa16,sa17,sa18,sa19,sa20,sa21,sa22,sa23,sa24,sa25,sa26,sa27,sa28,sa29,sa30,500,501,502,503,504,505,506,507}. Building upon the above concepts, we organize the article as follows: In Section II, we provide the field equations in F(R)-Euler-Heisenberg theory. In Section III, we obtain the black hole solutions in F(R)-Euler-Heisenberg gravity’s rainbow. In Sections IV and V, we study the thermodynamics, thermodynamic topology, and photon sphere of the mentioned black holes. We obtain the thermodynamic topology and topological classification of black holes in F(R)-Euler-Heisenberg gravity’s rainbow and its photon sphere in some cases in Section VI. Finally, we present the conclusion of our paper in Section VII.

\section{Field equations in $F(R)$-Euler-Heisenberg theory}
In this section, we aim to highlight the coupling of the EH term as a matter source with the $F(R)$ gravity. To this end, we consider the $F(R)$-EH action
\begin{equation}
    \mathcal{I}_{F(R)}=\frac{1}{16\pi}\int_{\mathcal{M}}\mathrm{d}^4x\sqrt{-g}\bigg(F(R)-\mathcal{L}(\mathcal{S},\mathcal{P})\bigg)
\end{equation}
where $F(R)=R+f(R)$,  in which $R$ and $f(R)$, are respectively, the Ricci scalar and a general function of the Ricci scalar, $g=\det (g_{\mu\nu})$ is the determinant of the metric tensor $g_{\mu\nu}$, and $\mathcal{L}(\mathcal{S},\mathcal{Q})$ is dedicated to being the EH Lagrangian. Throughout this study, we adopt the Newtonian gravitational constant and the speed of light as being equal to 1, i.e., $G=c=1$. In practical terms, the EH Lagrangian can be described by considering the following expression \cite{Heisenberg:1936nmg}:
\begin{equation}
    \mathcal{L}(\mathcal{S},\mathcal{Q})=-\mathcal{S}+\frac{\lambda}{2}\mathcal{S}^2+\frac{7\lambda}{8}\mathcal{Q}^2
\end{equation}
where $\lambda=8\alpha^2/45m_e^4$ is the EH parameter by which the intensity of the NLED contribution is regulated; $\alpha$ is the fine structure constant; and $m_e$ is the mass of the electron; hence, the EH parameter $(\lambda)$ is of the order of $\alpha/E^2_c$. In turn, $\mathcal{S}$ and $\mathcal{Q}$ are, respectively, true scalars and pseudo-scalars, with the following definitions:
\begin{equation}
    \mathcal{S}=\frac{\mathcal{F}}{4},\quad \mathcal{Q}=\frac{\Tilde{\mathcal{F}}}{4}
\end{equation}
where $\mathcal{F}=F_{\mu\nu}F^{\mu\nu}$ stands for the Maxwell invariant together with $F_{\mu\nu}=\partial_\mu A_\nu-\partial_\nu A_\mu$ is the electromagnetic field strength of which $A_\mu$ is the gauge potential. Furthermore, the invariant $\Tilde{\mathcal{F}}$ can be defined in terms of $\Tilde{\mathcal{F}}=F_{\mu\nu}\Tilde{F}^{\mu\nu}$, where $\Tilde{F}^{\mu\nu}=1/2\epsilon_{\mu\nu}^{\,\,\,\,\,\,\,\, \rho\lambda}F_ {\rho\lambda}$. Notably, the completely antisymmetric tensor $\epsilon_{\mu\nu\rho\sigma}$, satisfies $\epsilon_{\mu\nu\rho\sigma} \epsilon^{\mu\nu\rho\sigma} = -4$. It is worth pointing out that the $\lambda=0$ criterion abolishes the non-linear electromagnetic character of the EH theory so that the linear electromagnetic field of the Maxwell theory arises $(\mathcal{L}(\mathcal{S})=-\mathcal{S})$.

As far as NLEDs are concerned, though, there are two potential frameworks: one is the usual framework (the $\mathcal{F}$-frame) in terms of the electromagnetic field tensor $F_{\mu\nu}$. The other frame is the $P$ frame, with the $P_{\mu\nu}$ tensor as the principal field of the definition
\begin{equation}
    P_{\mu\nu}=-\left(\mathcal{L}_\mathcal{S} F_{\mu\nu}+\Tilde{F}_{\mu\nu}\mathcal{L}_\mathcal{P}\right)
\end{equation}
where $\mathcal{L}_{\mathcal{X}}=\frac{\partial\mathcal{L}}{\partial\mathcal{X}}$ with $X=(\mathcal{S}, \mathcal{Q})$. In the realm of the EH theory, $P_{\mu\nu}$ can be defined by
\begin{equation}
    P_{\mu\nu}=(1-\lambda F)F_{\mu\nu}-\Tilde{F}_{\mu\nu}\frac{7\lambda}{4}\mathcal{Q}\label{pp}
\end{equation}
which stands for a class of electric/magnetic fields, namely the electric induction $\mathbf{D}$ and the magnetic field $\mathbf{H}$. In connection to the condensed matter fields, the EH term, according to Eq. \eqref{pp}, involves a duality between the vectors set such that $\mathbf{D}$, $\mathbf{H}$ and the magnetic intensity $\mathbf{B}$ and the electric field $\mathbf{E}$.

To better grasp the EH model by means of the $P$ frame, it is quite worthwhile to consider the two independent invariants $P$ and $O$ assigned to the $P$ frame, which are given by
\begin{equation}
    P=-\frac{1}{4}P_{\mu\nu}P^{\mu\nu},\quad
 O=-\frac{1}{4}P_{\mu\nu}\,^*P^{\mu\nu}
\end{equation}
where $\Tilde{P}_{\mu\nu}=\frac{1}{2\sqrt{-g}}\epsilon_{\mu\nu\rho\sigma}P^{\sigma\rho}$. In particular, the Legendre transformation of the relevant Lagrangian $`\mathcal{L}$ is exploited to provide a structural function $\mathcal{H}$ expressed as follows:
\begin{equation}
    \mathcal{H}(P, O)=-\frac{1}{2}P^{\mu\nu}F_{\mu\nu}-\mathcal{L}
\end{equation}
         where, regardless of the second and higher order terms in $\lambda$, this structural function can be represented as follows:
\begin{equation}
    \mathcal{H}(P, O)=P-\frac{\lambda
    }{2}P^2-\frac{7\lambda}{8}O^2.
\end{equation}

The equations of motion of the $F(R)$-EH theory of gravity can be obtained, leading to
\begin{align}
  8\pi T_{\mu\nu}&= R_{\mu\nu}(1+f_R)-\frac{g_{\mu\nu}F(R)}{2}+(g_{\mu\nu}\nabla^2-\nabla_\mu\nabla_\nu)f_R\label{eom}\\
    \nabla_{\mu}P^{\mu\nu} &=0
\end{align}
where $f_R=\frac{\mathrm{d}f(R)}{\mathrm{d}R}$ and $T_{\mu\nu}$ is the energy-momentum tensor for the EH theory in the $P$ setting, given by
\begin{equation}
    T_{\mu\nu}=\frac{1}{4\pi}\bigg((1-\lambda P)P_\mu^\beta P_{\nu\beta}+g_{\mu\nu}(P-\frac{3}{2}\lambda P^2-\frac{7\lambda}{8}O^2)\bigg).
\end{equation}

Next, we present the spherically symmetric static solution for the $F(R)$-EH gravity’s rainbow.
\section{Black hole solutions in $F(R)$-Euler-Heisenberg gravity's rainbow}
Within the scope of this study, we consider a spherically, static, four-dimensional, energy-dependent spacetime, in which we adopt the procedure described in Refs. \cite{Magueijo:2002xx,peng2008covariant}, so that
\begin{equation}
h(\epsilon)=\eta^{\mu\nu}e_\mu(\epsilon)\otimes e_\nu(\epsilon)
\end{equation}
and
\begin{equation}
    e_0(\epsilon)=\frac{1}{f_{\epsilon}}\Tilde{e}_0,\qquad e_i(\epsilon)=\frac{1}{g_{\epsilon}}\Tilde{e}_i
\end{equation}
where the quantities in tilde (i.e. $\Tilde{e}_0$ and $\Tilde{e}_i$ ) stand for energy-independent frame fields. Based upon the foregoing assumptions, we in fact may generate black holes in $F(R)$-EH gravity’s rainbow by considering the following static, spherically symmetric, energy-dependent, four-dimensional spacetime
\begin{equation}
    \mathrm{d}s^2=-\frac{F(r)}{f^2_{\epsilon}}\,\mathrm{d}t^2+\frac{1}{g^2_{\epsilon}}\left(\frac{\mathrm{d}r^2}{F(r)}+r^2(\mathrm{d}\theta^2+\sin^2\theta\mathrm{d}\phi^2)\label{met}\right)
 \end{equation}
where $F(r)$, $f_{\epsilon}$ and $g_{\epsilon}$ is the metric function , and rainbow functions.

Practically speaking, the field equations underlying the context of $F(R)$ gravity with the EH matter fields (\ref{eom}) seem to be difficult. This means that accurate analytical solutions are proving tricky to pin down. To overcome this issue, one should consider the traceless energy-momentum tensor for the EH matter field, and then one can obtain an analytical solution from $F(R)$ gravity coupled to a nonlinear electrodynamics field. Thus, to be able to derive the solution of a black hole with constant curvature in the $F(R)$ theory of gravity as coupled to the EH matter field, it suffices for the trace of the stress-energy tensor $T_{\mu\nu}$ to hold to zero \cite{delaCruz-Dombriz:2009pzc,Moon:2011hq}. Based on the hypothesis of constant scalar curvature $R=R_0=\text{constant}$ \cite{Cognola:2005de}, the trace of equation \eqref{eom} takes the following terms:
\begin{equation}
R_0(1+f_{R_0})-2(R_0+f(_0))=0\label{tt}
\end{equation}
where $f_{R_0}=f_{R_{\mid R=R_0}}$. The constant scalar curvature is accordingly drawn from the equation \eqref{tt} as follows:
\begin{equation}
    R_0=\frac{2f(R_0)}{f_{R_0}-1 }.\label{uu}
\end{equation}

Making use of Eq. \eqref{uu} into Eq. \eqref{eom}, the field equations of the $F(R)$-EH theory could be established in the right way as
\begin{equation}
    R_{\mu\nu}(1+f_{R_0})-\frac{g_{\mu\nu} }{4}R_0(1+f_{R_0})=8\pi T_{\mu\nu}.\label{moe}
\end{equation}
By way of observation, the parameter $f_{R_0}$ implies a gravitational redefinition concerning general relativity; in particular, imposing the condition $f_{R_0}=0$ leads to a redefinition of the general relativity. Thus, the field equation in the $F(R)$-EH theory \eqref{moe} reduces to the EH theory of gravity.

Turning to the electromagnetic field, we have to treat only the electric field source of the EH theory so that the electromagnetic tensor is explicitly derived as
\begin{equation}
    P_{\mu\nu}=\frac{q}{r^2}\delta^0_{[\mu}\delta^1_{\nu]}\label{pp}
\end{equation}
in which the electromagnetic invariants are given by
\begin{equation}
    P=\frac{q^2}{2r^4},\quad O=0.
\end{equation}
where $q$ represents an integration constant that is associated with the electric charge.

At this stage, the field equations for $F(R)$ gravity with EH matter fields \eqref{eom}, in conjunction with the consideration of the metric function \eqref{met} and the tensor of the electric component \eqref{pp}, are explicitly expressed by the following differential equations:
    \begin{align}
    eq_{tt}= eq_{rr}=&rF'(r)+F(r)+\frac{r^2 R_0}{4g^2_{\epsilon}}-1\nonumber\\
    -&\frac{f^2_{\epsilon}}{(1+f_{R_0})}\left(\frac{\lambda q^2}{4r^6}-\frac{q^2}{r^2}\right),\\
    eq_{\theta\theta}= eq_{\phi\phi}=&rF''(r)+2F'(r)+\frac{r R_0}{2g^2_{\epsilon}}\nonumber\\
    +&\frac{f^2_{\epsilon}}{(1+f_{R_0})}\left(\frac{3}{4}\frac{\lambda q^4}{r^7}-\frac{q^2}{r^3}\right)
\end{align}

\begin{figure}[h!]
 \begin{center}
 \subfigure[]{
 \includegraphics[height=6cm,width=7cm]{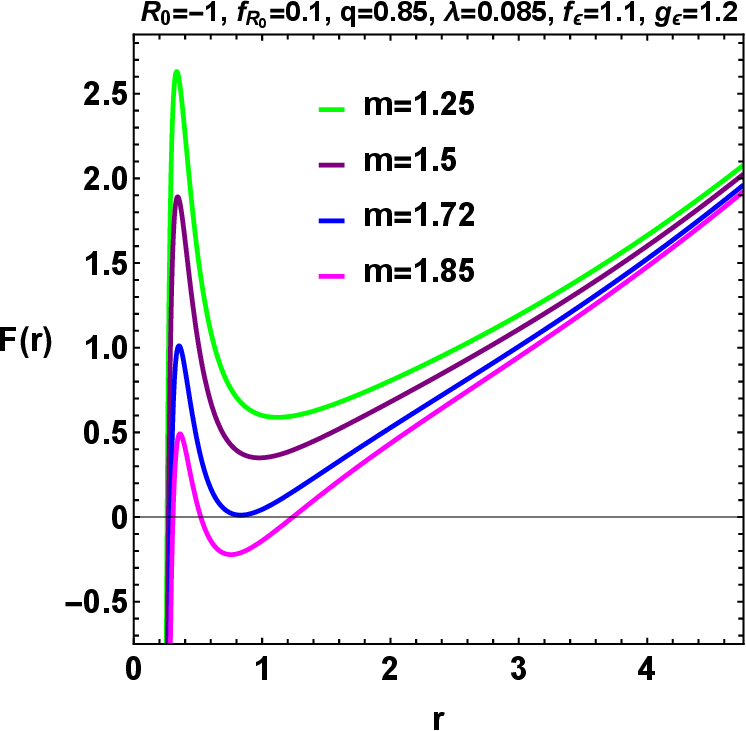}
 \label{1a}}
 \subfigure[]{
 \includegraphics[height=6cm,width=7cm]{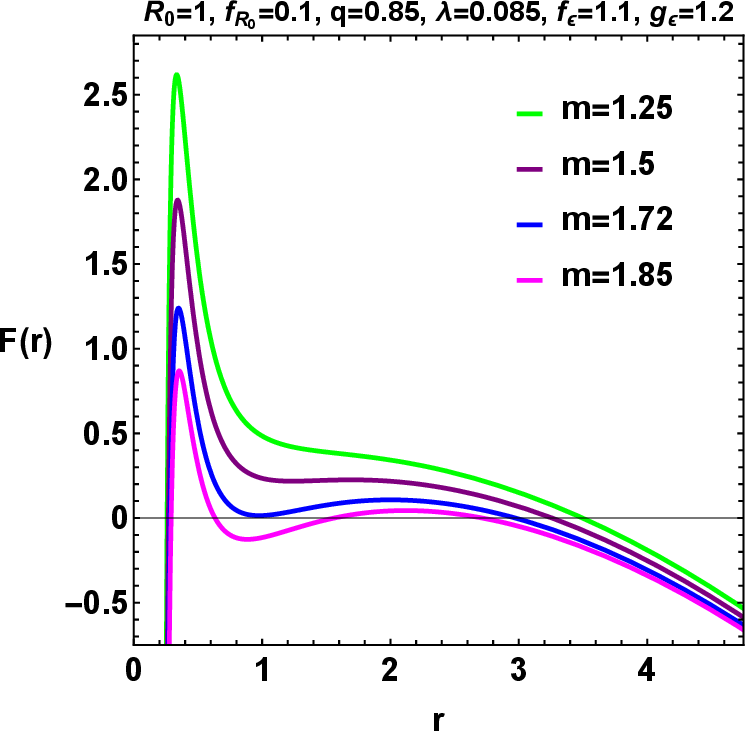}
 \label{1b}}
  \caption{\small{$F(r)$ versus $r$ for various values of the parameter $m_0$.}}
 \label{f1}
 \end{center}
 \end{figure}

\begin{figure}[h!]
 \begin{center}
 \subfigure[]{
 \includegraphics[height=5cm,width=5cm]{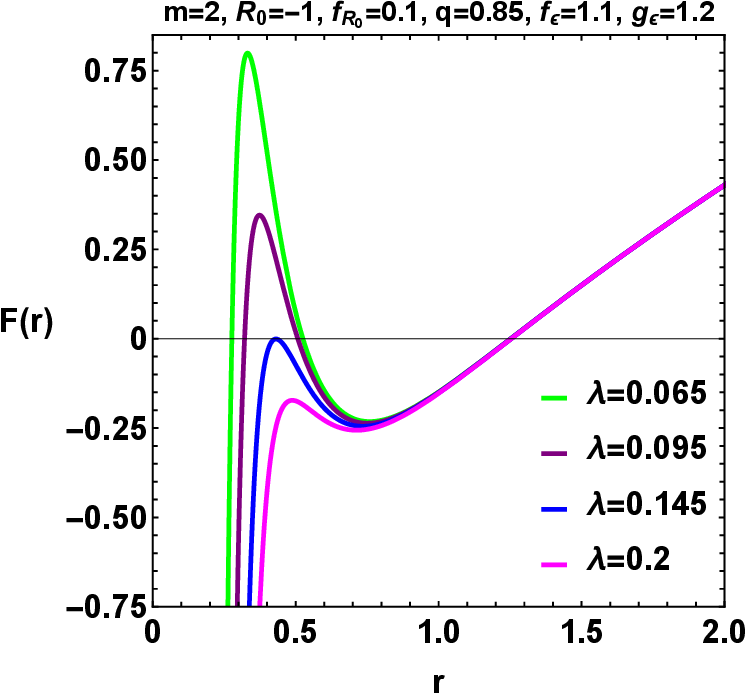}
 \label{ch1}}
 \subfigure[]{
 \includegraphics[height=5cm,width=5cm]{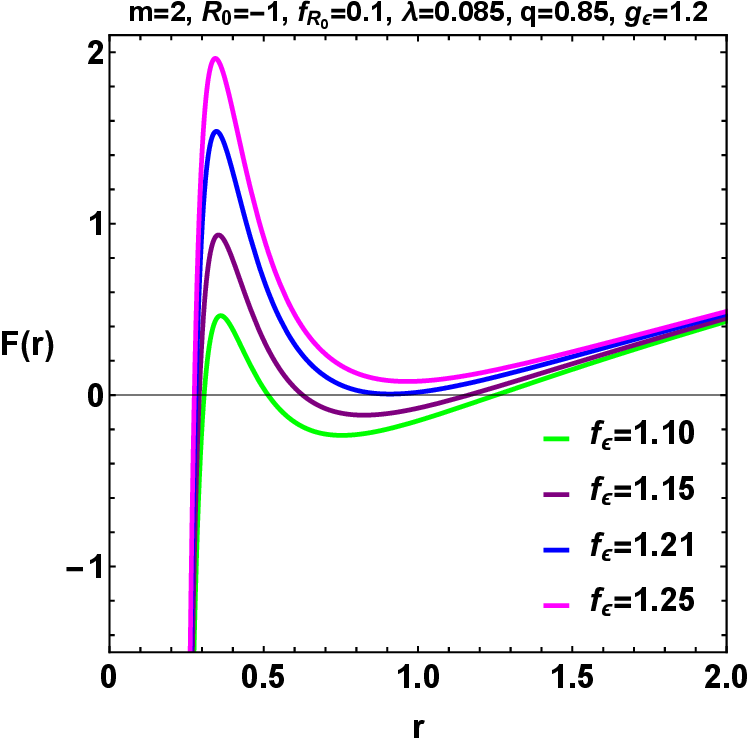}
 \label{ch2}}
 \subfigure[]{
 \includegraphics[height=5cm,width=5cm]{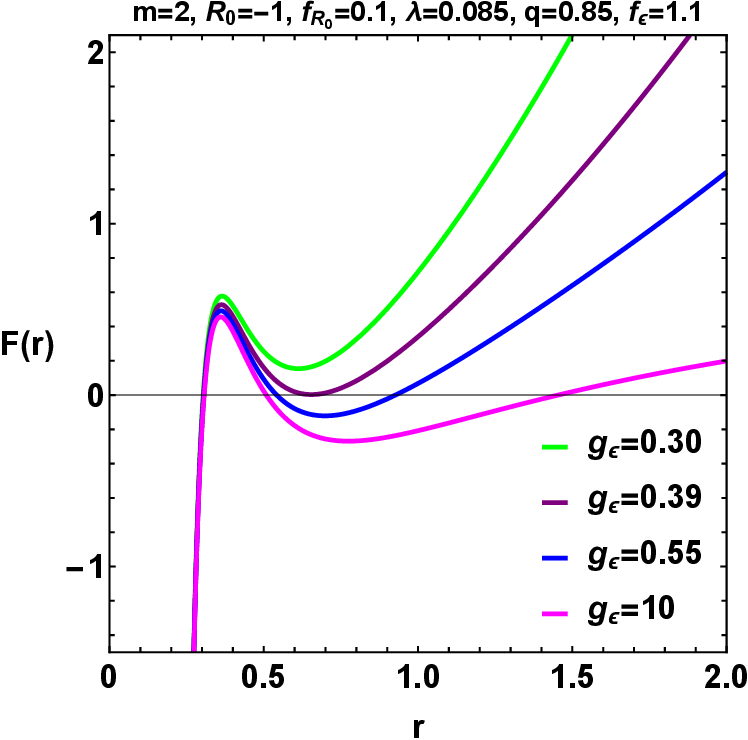}
 \label{ch3}}\\
 \subfigure[]{
 \includegraphics[height=5cm,width=5cm]{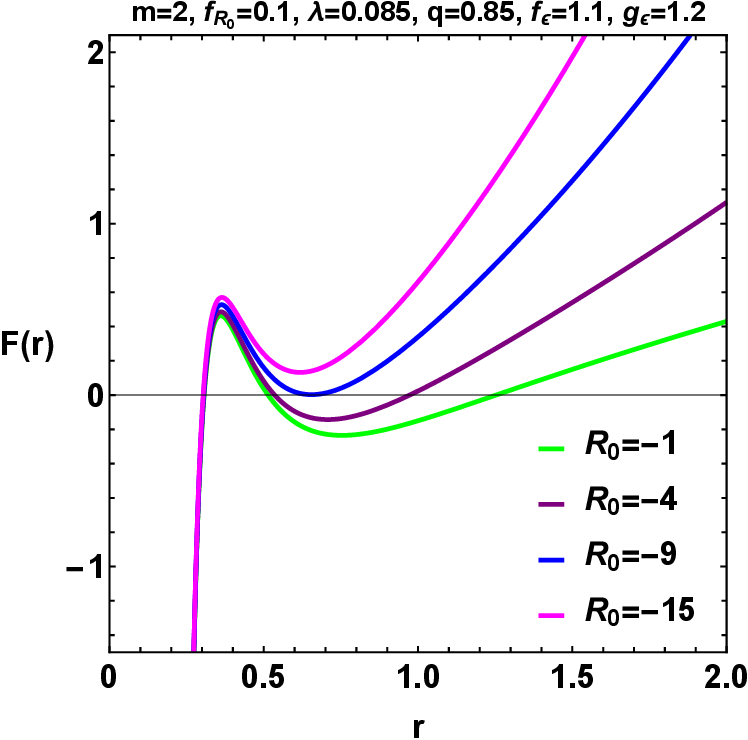}
 \label{ch4}}
 \subfigure[]{
 \includegraphics[height=5cm,width=5cm]{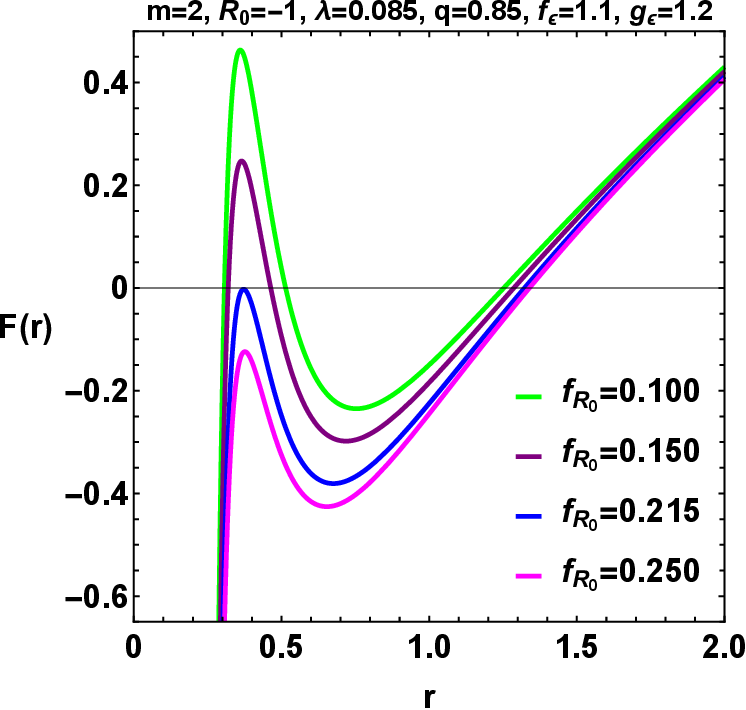}
 \label{ch5}}
 \subfigure[]{
 \includegraphics[height=5cm,width=5cm]{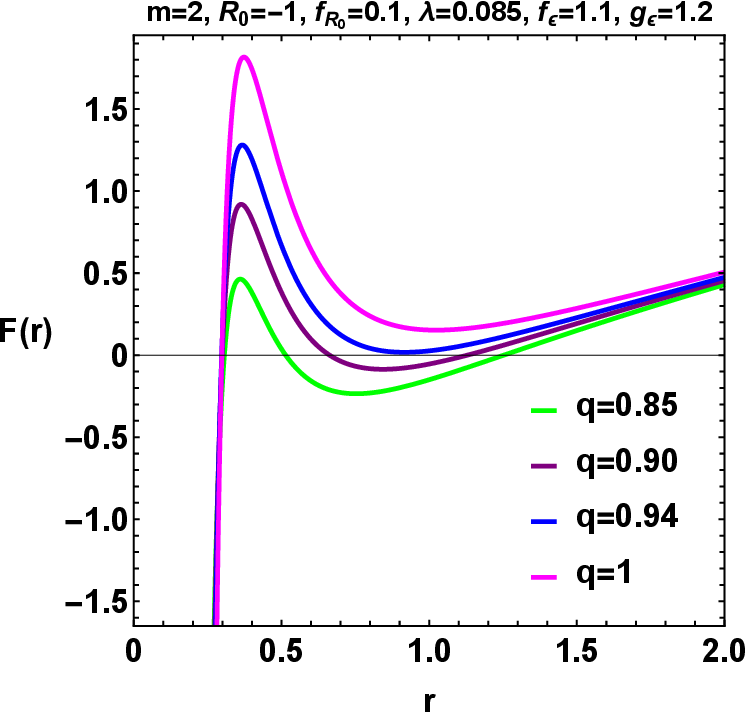}
 \label{ch6}}
  \caption{\small{$F(r)$ versus $r$ for various values of the parameter space $\left(R_0, f_{R_0}, q, \lambda, f_\epsilon, g_\epsilon\right).$}}
 \label{f2}
 \end{center}
 \end{figure}
where $eq_{tt}$, $eq_{rr}$, $eq_{\theta\theta}$ and $eq_{\phi\phi}$, are, respectively, the $tt$, $rr$, $\theta\theta$ and $\phi\phi$ components of the field equations \eqref{eom}. By implementing the above differential equation set, a definite solution to the constant scalar curvature $(R = R_0= \text{constant})$ can be provided. Following a close examination and further calculations, the metric function can now be expressed in the most precise form.

\begin{equation}
   F(r)=1-\frac{m}{r}-\frac{R_0 r^2}{12g^2_{\epsilon}} +\frac{f^2_{\epsilon}}{(1+f_{R_0}) }\bigg(\frac{q^2}{r^2}-\frac{\lambda q^4}{20 r^6}\bigg)\label{sol}
\end{equation}
where $m_0$ is an integration constant correlated with the black hole's geometric mass. It is worth highlighting that the obtained black hole solution \eqref{sol} is exactly in conformity with the field equations \eqref{eom}. Therefore, to carry out a physical analysis based on the black hole solution \eqref{sol}, one can ensure a constraint on the parameter $f_{R_0}$ such that $f_{R_0} \neq 1$. On the other hand, the consideration of the particular partial space, namely $(f_{R_0}=0, R_0=4\Lambda, f^2_{\epsilon}=g^2_{\epsilon}=1$ and $\lambda=0$), recovers the so-called Reissner-Nordstrom-(A)dS black hole defined by
\begin{equation}
   F(r)=1-\frac{m}{r}-\frac{\Lambda r^2}{3} +\frac{q^2}{r^2}-\frac{\lambda q^4}{20 r^6}.
\end{equation}
Such application tools are needed to understand the black hole solution better. In particular, the singularity and uniqueness features can be exploited by using the scalar invariants, i.e., the Ricci scalar, the Ricci square, and the Kretschmann scalar.

The Ricci scalar for the corresponding metric is given by
\begin{equation}\label{R}
    R=\frac{R_0}{g^2_{\epsilon}}+\frac{f^2_{\epsilon} \lambda  q^4}{\left(1+f_{R_0}\right)r^8 }.
\end{equation}
The Ricci squared is given by
    \begin{align}\label{RR}
 R_{\mu\nu}R^{\mu\nu}& = \frac{1}{4 g^4_{\epsilon}r^{16} \left(f_{R_0}+1\right)^2}\Bigg(5 f^4_{\epsilon} g^4_{\epsilon} \lambda ^2 q^8\nonumber\\
 &-16 f^4_{\epsilon} g^4_{\epsilon} \lambda  q^6 r^4+2 f^2_{\epsilon} g^2_{\epsilon} q^4 r^8 \nonumber\\
 &\times\bigg(8 f^2_{\epsilon}
   g^2_{\epsilon}+\lambda  R_0 \left(f_{R_0}+1\right)\bigg)\nonumber\\
   &+r^{16} R_0^2
   \left(f_{R_0}+1\right)^2\Bigg).
\end{align}
On the other hand, the Kretschmann scalar is formulated according to the metric function $\eqref{met}$ and the $F(r)$ fucntion (\ref{sol}) as follows:
    \begin{align}\label{RRR}
R_{\alpha\beta\mu\nu}R^{\alpha\beta\mu\nu}& = \frac{1}{150 g^4 r^{16}
   \left(f_{R_0}+1\right)^2}\Bigg(10 f^2 g^2 \lambda  q^4 r^4\nonumber\\
   &\times\bigg(r \left(f_{R_0}+1\right) \left(168 g^2 m_0+5 r^3
   R_0\right)\nonumber\\
   &-456 f^2 g^2 q^2\bigg)+25 r^8 \bigg(-288 f^2 g^4 m_0 q^2 r\nonumber\\
  &\times \left(f_{R_0}+1\right)+336 f^4 g^4 q^4+r^2 \left(f_{R_0}+1\right)^2\nonumber\\
  &\times\left(72 g^4
   m_0^2+r^6 R_0^2\right)\bigg)+717 f^4 g^4 \lambda ^2 q^8\Bigg).
\end{align}
Upon thorough analysis of the expressions \eqref{R}, \eqref{RR}, and \eqref{RRR}, the black hole solution described by the metric \eqref{met} is singular for all the allowable values of the parameters black hole system. So investigating the limits for the scalar invariants at $r = 0$  would result in the following:
\begin{align}
   \lim\limits_{r\to 0} \begin{cases}
    R\approx \infty
     \vspace{3mm}\\ R_{\mu\nu}R^{\mu\nu}\approx\infty
     \vspace{3mm}\\
R_{\alpha\beta\mu\nu}R^{\alpha\beta\mu\nu}\approx\infty.
    \end{cases}\,
\end{align}
By way of observation, the presence of the spacetime singularity is due to the mass $(m_0)$, the electric charge $(q)$, and the EH parameter $(\lambda)$ in the black hole metric. A useful way of avoiding the singularity resulting from the mass, the charge, and the EH parameters is to outline a procedure with a non-linear charge distribution function similar to Ref.~\cite{Balart:2014cga}. Throughout this work, we will not be thinking about such a situation and will stick to the metric function \eqref{met} for the rest of the analysis. On the other hand, examining behavior at large distances is a worthwhile adjunct, given that
\begin{align}
   \lim\limits_{r\to \infty} \begin{cases}
    R\approx \frac{R_0}{g^2_{\epsilon}}
     \vspace{3mm}\\ R_{\mu\nu}R^{\mu\nu}\approx\frac{R_0^2}{4g^4_{\epsilon}}
     \vspace{3mm}\\
R_{\alpha\beta\mu\nu}R^{\alpha\beta\mu\nu}\approx\frac{R_0^2}{4g^2_{\epsilon}}
    \end{cases}\,.
\end{align}
which implies that the Ricci scalar, the Ricci squared, and the Kretschmann scalar have a finite term at a large distance. To sum up, these scalars are evidence that our black hole solution is unique, and that the EH parameter  does not
modify the asymptotical behavior of the spacetime

Next, the examination of the metric function allows to explore the set of corresponding real roots. On the other hand, the analysis of the appropriate set of these real roots offers the possibility to extract information on the horizons (inner and outer horizons, and even more the cosmological horizon). Thus, to proceed with this examination, such a numerical approach is needed.

Solving the equation $B(r=r_h)=0$ at the event horizon leads to specifying the horizon radii spectrum. For this reason, Fig. \ref{f1} represents, either for $R_0<0$ (or AdS space if $R_0 = 4\Lambda$) or for $R_0>0$ (or dS space if $R_0 = 4\Lambda$), the variation of the metric function as a function of the space-time variable $r$ graphically showing the proper horizon radii. It is worth remarking that in the case of $R_0>0$, the horizon structure provides at most four horizon radii (one of which refers to the cosmological horizon). In contrast, as we can observe, the case related to $R_0<0$ can cover only three possible horizon radii (one of which is adapted to the cosmological horizon). Roughly speaking, the sign of the parameter $R_0$ plays a crucial role in defining the  structure of black hole horizon in the $F(R)$-EH gravity’s rainbow.

Analysing the impact of the electrical charge ($q$), the parameter of EH ($\lambda$), and $F(R)$’s
parameters ($f_{R_0}$, and $R_0$) and Rainbow structure ($f_{\epsilon}$, and $g_{\epsilon}$) on the structure of the black hole horizon in the $F(R)$-EH gravity’s rainbow is also necessary. To this end, Fig. \ref{f2} depicts the behaviour of the metric function against the spacetime variable $r$. It is observed that the appropriate horizon structure provides at most three horizon radii (one of which corresponds to the cosmological horizon) for the case where $R_0<0$. Thus, each parameter variation affects the horizon structure in such a way that we can point out that at a critical value of the EH parameter, i.e., $\lambda_{\text{crit}}=0.145$, there are two horizon radii, which are a root (in the extreme case) and the cosmological horizon. Further, at the range $\lambda_{crit}<\lambda$, the horizon structure admits three horizon radii, namely the inner horizon, the outer horizon, and the cosmological horizon. Whereas for $\lambda_{crit}>\lambda$, the horizon structure merely carries the cosmological horizon. Similarly, a graphical inspection of the electrical charge parameter $(q)$ may imply a particular pattern for the horizon structure. So, we can observe that for a certain critical value, such as $q_{\text{crit}}=0.95$, the horizon structure may entail two horizon radii, the inner horizon being the smallest among the three horizons and the other being formed by a matching between the outer horizon and the cosmological horizon. Furthermore, for $q_{crit}>q$, the horizon structure involves only the inner horizon, whereas for $q_{crit}<q$, the situation consists of three horizon radii. As far as the parameters of $F(R)$ are concerned, namely $f_{R_0}$ and $R_0$, the horizon structure, for instance, in terms of $R_0$ entails a critical value $R_0^{\text{crit}}=-9.45$ at which it exhibits two types of horizon, the inner horizon and the other being formed by a correlation between the outer horizon and the cosmological horizon. In addition, for $R^{crit}_0>R_0$, the black hole horizon carries three horizon radii, while in the case of $R^{crit}_0<R_0$, the horizon structure has only one horizon, which stands for the outer. On the other hand, it is notable that for a critical value of the parameter $f_{R_0}$ such that $f_{R_0}^{\text{crit}}=0.215$, the black hole admits in particular the extreme horizon and the cosmological horizon. For $f_{R^{crit}_0}>f_{R_0}$, the situation turns out to be different since the black hole's horizon entails only the cosmological horizon. Moreover, in the case $f_{R^{crit}_0}<f_{R_0}$, the horizon structure exhibits three horizon radii, i.e., all the possible horizons of the black hole. Finally, it is necessary to explore the characteristic of the horizon structure concerning the variation of the set of rainbow gravity functions, such as ($f_{\epsilon}$, $g_{\epsilon}$). This variation can be appreciated on the plot, in which, at a critical value of $g_{\epsilon}$ as $g_{\epsilon}^{\text{crit}}=0.39$, the horizon structure possesses the inner horizon and one that is correlated between the outer horizon and the cosmological horizon. Other scenarios can be considered to clearly assess the impact of the Rainbow function such as $g_{\epsilon}$ on the horizon, from which at $g_{\epsilon}> g_{\epsilon}^{\text{crit}}$, the black hole admits three horizons, whereas for $g_{\epsilon}< g_{\epsilon}^{\text{crit}}$, the horizon involves only the inner horizon. On the other hand, we observe that the horizon structure comprises the inner horizon and a correlated horizon between the outer horizon and the cosmological horizon in the case where $f_{\epsilon}^{\text{crit}}=1.21$, i.e., above this threshold, the black hole admits only the inner horizon, and below this threshold, the black hole presents three horizon radii.

\section{Thermodynamics}
\label{sec5}
The best way to approach thermodynamically the black hole solutions in $F(R)$ gravity’s rainbow with EH matter source is to consider the first law of thermodynamics. To begin with, inspection at the event horizon $r=r_h$ of the metric solution \eqref{sol} leads to providing the mass term of the black hole as follows:
\begin{equation}\label{mass}
    m=r_h-\frac{r_h^3 R_0}{12 g^2_{\epsilon}}-\frac{f^2_{\epsilon}}{\left(f_{R_0}+1\right)}\left(\frac{ \lambda  q^4}{20 r_h^5 }+\frac{q^2}{r}\right).
    \end{equation}
    Next, to ascertain the Hawking temperature, one has to consider primarily the surface gravity~\cite{Kubiznak:2016qmn}, which is supplied out of
\begin{equation}\label{surface}
    \kappa=\left(-\frac{1}{2}\nabla_\mu\xi_\nu\nabla^\mu\xi^\nu\right)^{1/2}\bigg\vert_{r=r_h}=\frac{F'(r)g_{\epsilon}}{f_{\epsilon}}\bigg\vert_{r=r_h}
\end{equation}
with $\xi^\mu=\partial/\partial t$ is a Killing vector. Achieving a correct expression of the surface gravity $(\kappa)$ is carried out by exploiting the metric function \eqref{sol} and by injecting the mass \eqref{mass} into Eq. \eqref{surface}. The surface gravity of a black hole is therefore expressed by
\begin{eqnarray}
  \kappa&=&\frac{f_{\epsilon} g_{\epsilon} }{8 \left(f_{R_0}+1\right)
   r_h^7}\left(\lambda  q^4-4 q^2 r_h^4\right)-\frac{R_0 r_h}{8 f_{\epsilon} g_{\epsilon}}\nonumber\\
   &+&\frac{g_{\epsilon}}{2 f_{\epsilon} r_h},
\end{eqnarray}
as well as the formula $T=\kappa/2 \pi$ stands for the Hawking temperature, which could be expressed by an appropriate term as follows:
\begin{eqnarray}\label{tem}
  T&=&\frac{f_{\epsilon} g_{\epsilon} }{16\pi \left(f_{R_0}+1\right)
   r_h^7}\left(\lambda  q^4-4 q^2 r_h^4\right)-\frac{R_0 r_h}{16\pi f_{\epsilon} g_{\epsilon}}\nonumber\\
   &+&\frac{g_{\epsilon}}{4\pi f_{\epsilon} r_h}.
\end{eqnarray}
It is noteworthy that all the parameters set in the black hole system, namely, $R_0, f_{R_0}$, $q$, $\lambda$, $g_{\epsilon}$ and $f_{\epsilon}$ perfectly affected the behavior of the Hawking temperature.

To achieve an accurate description of the Hawking temperature behavior, it is merely instructive to study the asymptotic and high-energy limits of the temperature. In this respect, the high-energy limit can be exemplified as follows
\begin{equation}
 \lim\limits_{r_h\to 0} T\propto\frac{f_{\epsilon} g_{\epsilon} \lambda  q^4}{16\pi \left(f_{R_0}+1\right)
   r_h^7}
\end{equation}
which proves the extent to which the parameters of the matter sector $(\lambda, q)$ as well as the rainbow functions $(f_{\epsilon},g_{\epsilon})$ influence the high-energy behavior of the Hawking temperature. Accordingly, the Hawking temperature is definitely positive for at the high-energy limit.

On the other hand, the asymptotic limit of temperature remains just a dependence of $R_0$, such that
\begin{equation}
 \lim\limits_{r_h\to \infty} T\propto-\frac{R_0 r_h}{16\pi f_{\epsilon} g_{\epsilon}}
\end{equation}
which implies that the Hawking temperature sets to be positive (negative) whenever $R_0<0$ ($R_0>0$).

By using Gauss's law, the electric charge of the black hole per unit volume, $V$, can be ascertained by the following formula
\begin{equation}
    Q=\frac{\Tilde{Q}}{V}=\frac{qf_{\epsilon}}{4\pi g_{\epsilon}}.
\end{equation}

The EH $\Phi$ electromagnetic potential corresponding to the black hole within the framework of the F(R)-EH gravity's rainbow is as follows
\begin{eqnarray}\label{charge}
\Phi&=&\int_{r_h}^\infty\mathrm{d}r P_{tr}\mathcal{H}_P(P,0)=\int_{r_h}^\infty\mathrm{d}r\frac{q}{r^2}\left(1-\frac{\lambda q^2}{2 r^4}\right)\nonumber\\
    &=&\frac{q}{r_h}\left(1-\frac{\lambda q^2}{10 r_h^4}\right).
\end{eqnarray}

By applying a modified area law in the $F(R)$ theory of gravity, we can extrapolate the corresponding entropy of black holes such as
\begin{eqnarray}\label{a1}
    S=\frac{\mathcal{A}(1+f_{R_0})}{4}
\end{eqnarray}
with $\mathcal{A}$ being the horizon area as
\begin{equation}\label{a2}
\mathcal{A}=\int_{0}^{2\pi}\int_0^\pi\sqrt{g_{\theta\theta}g_{\phi\phi}}\bigg\vert_{r=r_h}=\frac{r^2}{g^2_{\epsilon}}\bigg\vert_{r=r_h}=\frac{r_h^2}{g^2_{\epsilon}}
\end{equation}
So now we can work out the entropy of EH black holes in $F(R)$ gravity's rainbow by substituting Eq. \eqref{a2} into Eq. \eqref{a1}, thus yielding
\begin{equation}
    S=\frac{\Tilde{S}}{\mathcal{V}}=\frac{(1+f_{R_0})r_h^2}{4g^2_{\epsilon}}
\end{equation}
which states that the law of areas cannot be held for black hole solutions in gravity $R + f(R)$.

Exploring the total mass of black holes in $F(R)$-EH theory would be achievable by looking at the Ashtekar-Magnon-Das (AMD) approach \cite{Ashtekar:1984zz,Ashtekar:1999jx}, which can be expressed as follows
\begin{equation}\label{ADM}
    M=\frac{\Tilde{M}}{\mathcal{V}}=\frac{m(1+f_{R_0})}{8\pi g_{\epsilon}f_{\epsilon}}.
\end{equation}
This allows us, after substituting mass \eqref{mass} in Eq. \eqref{ADM}, to obtain the total mass as follows:
\begin{equation}\label{M}
    M=\frac{\left(f_{R_0}+1\right) \left(12 g_{\epsilon}^2 r_h-R_0 r_h^3\right)}{96 \pi  f_{\epsilon} g^3_{\epsilon}}+\frac{f_{\epsilon}
   \left(20 q^2 r_h^4-\lambda  q^4\right)}{160 \pi  g_{\epsilon} r_h^5}
\end{equation}

Afterwards, inspection of the total mass at its high-energy limit can reveal the form
\begin{equation}
 \lim\limits_{r_h\to 0} M\propto-\frac{f_{\epsilon}
   \lambda  q^4}{160 \pi  g_{\epsilon} r_h^5}
\end{equation}
where a consistent dependency is shown only with parameters $(q,\lambda, g_{\epsilon}$ and $f_{\epsilon}$). Thus, at the high-energy limit, it is striking that the total mass of small black holes is always negative.

The asymptotic limit, on the other hand, of the total mass is found to be as follows:
\begin{equation}
 \lim\limits_{r_h\to \infty} M\propto-\frac{\left(f_{R_0}+1\right) R_0 r_h^3}{96 \pi  f_{\epsilon} g^3_{\epsilon}}
\end{equation}
where the positive (negative) branch of $M$ is consistently referred to as $R_0<0$ ($R_0>0$).

At this stage, we may evaluate the first law of thermodynamics.  Accordingly, the conserved and specific thermodynamic quantities in Eqs. \eqref{M}, \eqref{tem}, \eqref{charge} and \eqref{a1} satisfy the first law of thermodynamics in the following format.
\begin{equation}
    \mathrm{d}M=T\mathrm{d}S+\Phi\mathrm{d}Q
\end{equation}
where $T=\left(\frac{\partial M}{\partial S}\right)_Q$, and $\Phi=\left(\frac{\partial M}{\partial Q}\right)_S$, respectively, are in accordance with the relations achieved in the Eqs. \eqref{tem} and \eqref{charge}.
\section{Thermodynamic topology and photon sphere}
\label{sec6}
Thermodynamic topology involves studying the thermodynamic properties of black holes using topological methods. It helps understand the critical points and phase transitions of black holes by analyzing their topological charges and numbers. A photon sphere is a region where gravity is strong enough that photons (light particles) are forced to travel in orbits. For a black hole, this sphere lies just outside the event horizon. The existence and properties of photon spheres can be studied using topological methods, which assign topological charges to these spheres. These charges help in understanding the stability and structure of the photon sphere. In recent studies, researchers have explored the relationship between thermodynamic topology and photon spheres in various black hole models, including those with hyperscaling violation. They have found that different topological charges correspond to different physical configurations, such as black holes and naked singularities\cite{sa16}. Here, we want to study the Thermodynamic topology of Black Holes in $F(R)$-Euler-Heisenberg gravity's Rainbow. so we follow some route that is explained in detail in the following section
\section{Thermodynamics topology of black holes}
The primary concept associated with defects is the topological charge. To examine the thermodynamic topology of a black hole, we calculate the topological charge and utilize it to determine the topological classes. We employ Duan’s $\phi$ mapping technique to compute the topological charge. To introduce the thermodynamic properties of black holes, various quantities are utilized. For example, two distinct variables, such as mass and temperature, can describe the generalized free energy\cite{Wei:2022dzw}. Given the relationship between mass and energy in black holes, we can reformulate our generalized free energy function as a standard thermodynamic function in the following form,
\begin{equation}\label{1'}
\mathcal{F}=M-\frac{S}{\tau},
\end{equation}
where, $\tau$ represents the Euclidean time period, and \( T \) (the inverse of $\tau$) denotes the temperature of the ensemble. The generalized free energy is on-shell only when $\tau = \tau_{H} = \frac{1}{T_{H}}$. From the expression for the off-shell free energy of a black hole, a vector field is constructed as follows,
\begin{equation}\label{2'}
\phi=(\phi^{r_H},\phi^{\Theta})=\big(\frac{\partial\mathcal{F}}{\partial r_{H}} ,-\cot\Theta\csc\Theta\big),
\end{equation}
where $\phi^{\Theta}$ is divergent, the vector direction points outward when $\Theta = 0$ or $\pi$. For $r_{H}$ and $\Theta$, the ranges are $0 \leq r_{H} \leq \infty$ and $0 \leq \Theta \leq \pi$, respectively. A topological current can be defined using Duan's $\phi$-mapping topological current theory as follows,
\begin{equation}\label{3'}
j^{\mu}=\frac{1}{2\pi}\varepsilon^{\mu\nu\rho}\varepsilon_{ab}\partial_{\nu}n^{a}\partial_{\rho}n^{b},\hspace{1cm}\mu,\nu,\rho=0,1,2
\end{equation}
To determine the topological charge, we first identify the unit vector \( \mathbf{n} \), where \( \mathbf{n} = (n^1, n^2) \). We have \( n^1 = \frac{\phi^{r_H}}{\|\phi\|} \) and \( n^2 = \frac{\phi^\Theta}{\|\phi\|} \). According to Noether's theorem, the resulting topological currents are conserved viz $\partial_{\mu}j^{\mu}=0,$. It can be demonstrated that the current \( j^\mu \) is non-zero only at the zero points of the vector field by utilizing the following equation,
\begin{equation}\label{4'}
j^{\mu}=\delta^{2}(\phi) J^{\mu}(\frac{\phi}{x}),
\end{equation}
where we utilize the following properties of the Jacobi tensor,
\begin{equation}\label{5'}
\varepsilon^{ab}J^{\mu}(\frac{\phi}{x})=\varepsilon^{\mu\nu\rho}\partial_{\nu}\phi^{a}\partial_{\rho}\phi^{b}.
\end{equation}
The Jacobi vector reduces to the standard Jacobi when \( \mu = 0 \), as demonstrated by \( J^{0}\left(\frac{\phi}{x}\right) = \frac{\partial(\phi^1, \phi^2)}{\partial(x^1, x^2)} \). Equation (\ref{4'}) shows that \( j^{\mu} \) is non-zero only when \( \phi = 0 \). Through some calculations, we can express the topological number or total charge \( W \) in the following form,
\begin{equation}\label{6'}
W=\int_{\Sigma}j^{0}d^2 x=\Sigma_{i=1}^{n}\beta_{i}\eta_{i}=\Sigma_{i=1}^{n}\omega_{i},
\end{equation}
where $\beta_i$ is the positive Hopf index, counting the loops of the vector $\phi^a$ in the $\phi$ space when $x^\mu$ is near the zero point $z_i$. Meanwhile, $\eta_i = \text{sign}(j^0(\phi/x)_{z_i}) = \pm 1$. The quantity $\omega_i$ represents the winding number for the $i$-th zero point of $\phi$ in $\Sigma$. Note that the winding number is independent of the shape of the region where the calculation occurs. The value of the winding number is directly associated with black hole stability. The topological charge \( W \) can be determined by summing the winding numbers calculated along each contour surrounding the zero points,
\begin{equation}\label{7'}
W=\sum_{i}\omega_{i}.
\end{equation}
It is important to note that if the parameter region lacks zero points, the total topological charge is considered to be zero. In the following two subsections, we will explore the thermodynamic topology of black holes in Euler-Heisenberg F(R)-Rainbow gravity
\subsection{Topological classification of black holes in $F(R)$-Euler-Heisenberg gravity's Rainbow}
\label{sec7}
\begin{figure}[h!]
 \begin{center}
 \subfigure[]{
 \includegraphics[height=2.6cm,width=2.6cm]{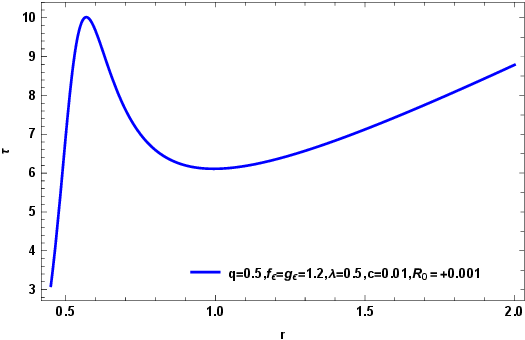}
 \label{3a}}
 \subfigure[]{
 \includegraphics[height=2.6cm,width=2.6cm]{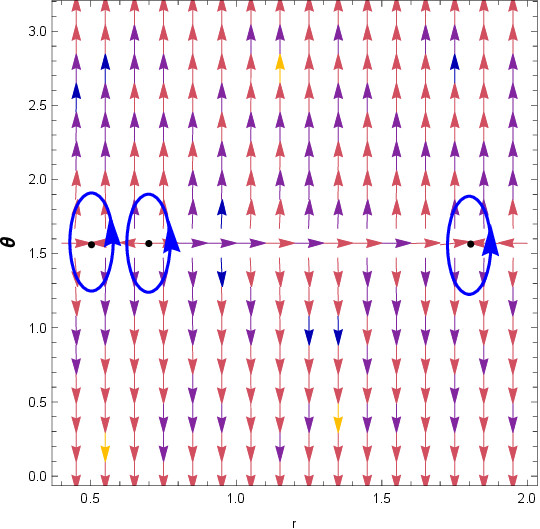}
 \label{3b}}
 \subfigure[]{
 \includegraphics[height=2.6cm,width=2.6cm]{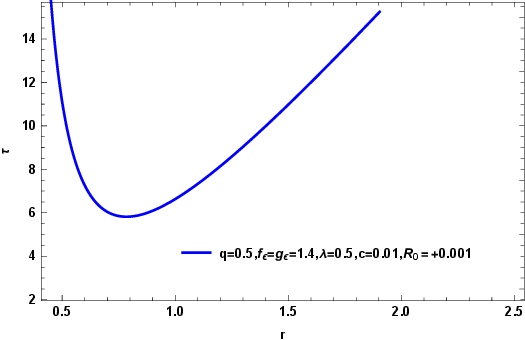}
 \label{3c}}
 \subfigure[]{
 \includegraphics[height=2.6cm,width=2.6cm]{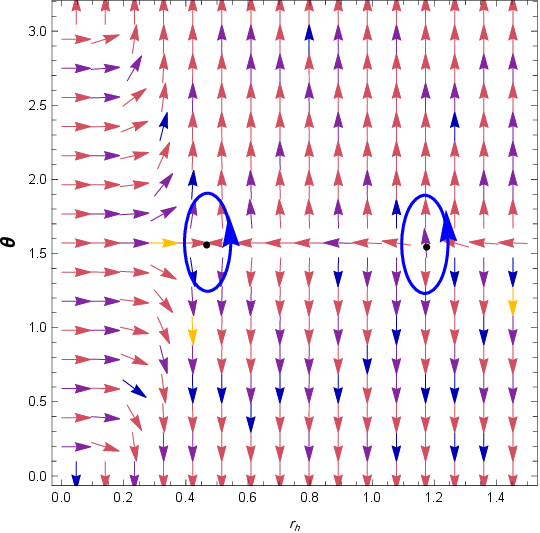}
 \label{3d}}
 \subfigure[]{
 \includegraphics[height=2.6cm,width=2.6cm]{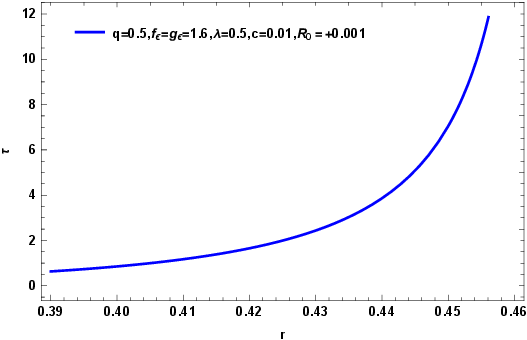}
 \label{3e}}
 \subfigure[]{
 \includegraphics[height=2.6cm,width=2.6cm]{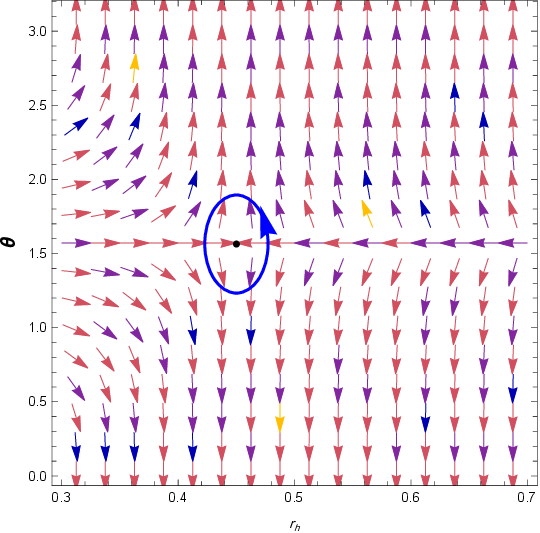}
 \label{3f}}
  \caption{\small{The arrows illustrate the vector field \( n \) on a segment of the \( (r_h-\theta) \) plane for black holes in Euler-Heisenberg F(R)-Rainbow gravity, with parameters \( q = 0.5 \), \( \lambda = 0.5 \), \( c = 0.01 \), \( R_0 = +0.001 \), and \( f_{\epsilon} = g_{\epsilon} = 1.2, 1.4, 1.6 \) respectively. The zero points (ZPs) are located within the circular loop at \( (r_h, \theta) \). The contours (blue loop) enclose the ZPs.}}
 \label{f3}
 \end{center}
 \end{figure}

\begin{figure}[h!]
 \begin{center}
 \subfigure[]{
 \includegraphics[height=2.6cm,width=2.6cm]{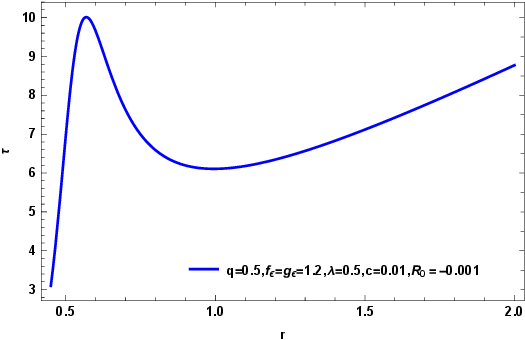}
 \label{4a}}
 \subfigure[]{
 \includegraphics[height=2.6cm,width=2.6cm]{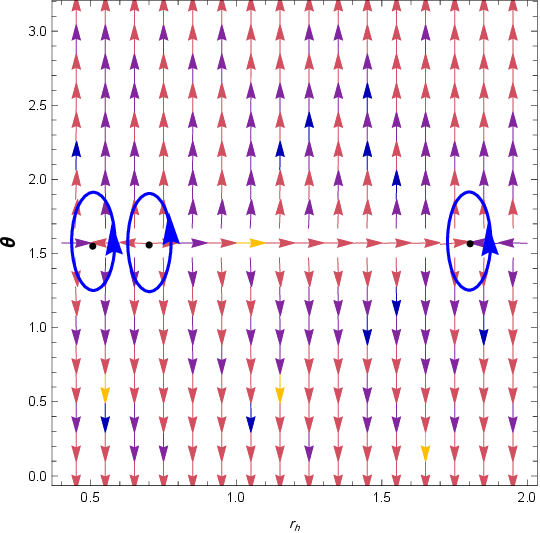}
 \label{4b}}
 \subfigure[]{
 \includegraphics[height=2.6cm,width=2.6cm]{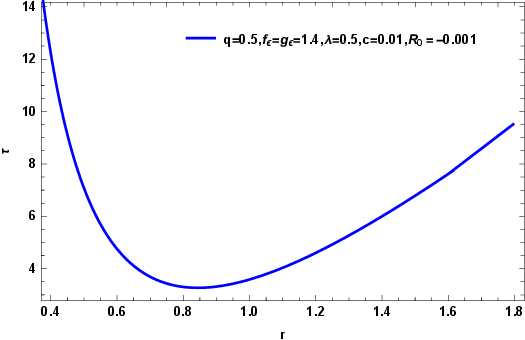}
 \label{4c}}
 \subfigure[]{
 \includegraphics[height=2.6cm,width=2.6cm]{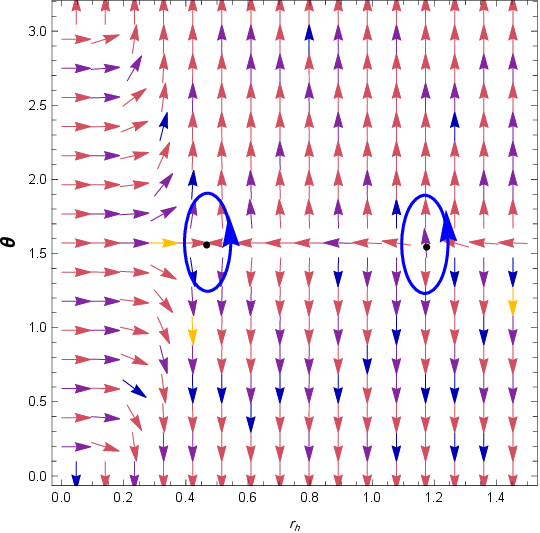}
 \label{4d}}
 \subfigure[]{
 \includegraphics[height=2.6cm,width=2.6cm]{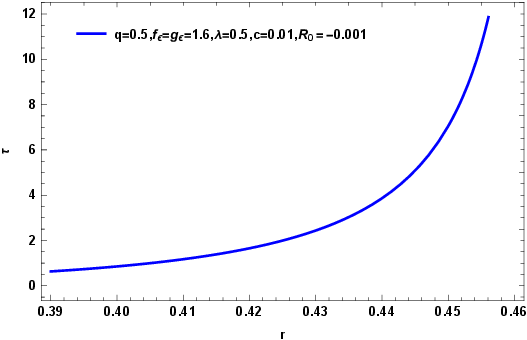}
 \label{4e}}
 \subfigure[]{
 \includegraphics[height=2.6cm,width=2.6cm]{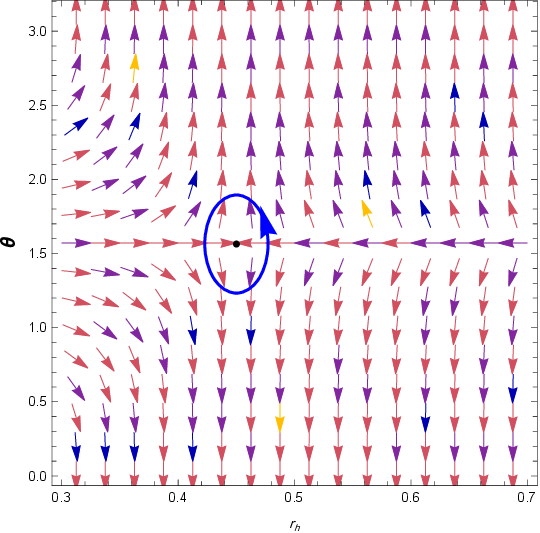}
 \label{4f}}
  \caption{\small{The arrows illustrate the vector field \( n \) on a segment of the \( (r_h-\theta) \) plane for black holes in Euler-Heisenberg F(R)-Rainbow gravity, with parameters \( q = 0.5 \), \( \lambda = 0.5 \), \( c = 0.01 \), \( R_0 = -0.001 \), and \( f_{\epsilon} = g_{\epsilon} = 1.2, 1.4, 1.6 \) respectively. The zero points (ZPs) are located within the circular loop at \( (r_h, \theta) \). The contours (blue loop) enclose the ZPs.}}
 \label{f4}
 \end{center}
 \end{figure}

Based on equations (\ref{1'}), we derive the generalized Helmholtz free energy for black holes within the framework of Euler-Heisenberg F(R)-Rainbow gravity
\begin{equation}\label{8'}
\begin{split}
\mathcal{F}=\frac{-120 (c+1) f_{\epsilon} g_{\epsilon} r_H^7+5 (c+1) r_H^6 \tau  \left(r_H^2 R_0-12 g_{\epsilon}^2\right)+3 f_{\epsilon}^2 g_{\epsilon}^2 q^2 \tau  \left(\lambda  q^2-20 r_H^4\right)}{480 \pi  f_{\epsilon} g_{\epsilon}^3 r_H^5 \tau}
\end{split}
\end{equation}
From the discussion in the previous section, the form of the function $(\phi^r,\phi^{\Theta} )$ is determined as follows,
\begin{equation}\label{9'}
\begin{split}
&\phi^{r_H}=\frac{-16 (c+1) f_{\epsilon}g_{\epsilon} r_H^7+(c+1) r_H^6 \tau  \left(4 g_{\epsilon}^2-r_H^2 R_0\right)+f_{\epsilon}^2 g_{\epsilon}^2 q^2 \tau  \left(\lambda  q^2-4 r_H^4\right)}{32 \pi  f_{\epsilon} g_{\epsilon}^3 r_H^6 \tau}\\
&\phi^{\Theta}=-\frac{\cot (\Theta )}{\sin (\Theta )}\\
\end{split}
\end{equation}
The unit vectors \( \mathbf{n}_1 \) and \( \mathbf{n}_2 \) are computed using equation (\ref{9'}). Next, we calculate the zero points of the \( \phi^{r_H} \) component by solving the equation \( \phi^{r_H} = 0 \) and derive an expression for \( \tau \) as follows,
\begin{equation}\label{10'}
\begin{split}
\tau=\frac{\big[16 (c+1) f_{\epsilon} g_{\epsilon} r_H^7\big]}{\big[4 c g_{\epsilon}^2 r_H^6-c r_H^8 R_0+f_{\epsilon}^2 g_{\epsilon}^2 \lambda  q^4-4 f_{\epsilon}^2 g_{\epsilon}^2 q^2 r_H^4+4 g_{\epsilon}^2 r_H^6-r_H^8 R_0{}\big]}
\end{split}
\end{equation}
We analyze Figure (\ref{f3}), and (\ref{f4}), which pertain to the structure of black holes in Euler-Heisenberg F(R)-Rainbow gravity. The figures display the normalized field lines. As depicted in the figures, some zero points represent the model's topological charges with various free parameters. They are proportional to the winding number and are situated inside the blue contour at the coordinates \((r_H, \Theta)\).
We have plotted Figures (\ref{f3}) for the mentioned model with respect to \( q = 0.5 \), \( \lambda = 0.5 \), \( c = 0.01 \), \( R_0 = +0.001 \) to draw these contours. In Figure (\ref{3b}), we plotted the curve for \( f_{\epsilon} = g_{\epsilon} = 1.2 \).
As shown in Figure (\ref{3b}), three blue loops represent the topological numbers $\omega$=(-1, +1, -1) and the total topological number \( W = -1 \). Also, in Figures (\ref{3d}) and (\ref{3f}), we plotted the curve for \( f_{\epsilon} = g_{\epsilon} = 1.2, 1.4 \) respectively. (\ref{3d}) represents the topological numbers \(-1, +1\) with the total topological number \( W = 0 \) and Figure (\ref{3f}) represents the one topological number \(-1\), so that the total topological number is \( W = -1 \).
Without loss of generality, we analyzed the topological properties of these black holes in this paper by choosing \(\tau\) for black holes in Euler-Heisenberg F(R)-Rainbow gravity. The content discusses the stability of the black holes by examining the winding numbers and specific heat capacity. We have plotted Figures (\ref{f4}), for the mentioned model with respect to free parameters \( q = 0.5 \), \( \lambda = 0.5 \), \( c = 0.01 \), \( R_0 = -0.001 \) and \( f_{\epsilon} = g_{\epsilon} = 1.2, 1.4, 1.6 \) respectively, following the same approach.
We plotted the normalized field vectors \( n \) for these black holes in Figures (\ref{f4}). As it is clear in these figures, for Figure (\ref{4b}) corresponding to the free parameters \( q = 0.5 \), \( \lambda = 0.5 \), \( c = 0.01 \), \( R_0 = -0.001 \) and \( f_{\epsilon} = g_{\epsilon} = 1.2 \), we have topological charges $\omega$=(-1, +1, -1) and the total topological charge \( W = -1 \). For the other two samples with \( f_{\epsilon} = g_{\epsilon} = 1.4, 1.6 \), the total topological charges are equal to $\omega$=(-1, +1) and $\omega$=(-1) respectively which are shown in  Figure (\ref{4d}) and Figure (\ref{4f}). By comparing Figures (\ref{f3}) and (\ref{f4}), it is clear that the positive or negative of \( R_0 \) and parameter values of \( f_{\epsilon} = g_{\epsilon} \) play a very important role in determining the topological charges. We summarize the results in Table I.

 \begin{table*}
\begin{tabularx}{\textwidth}{X}
  \centering
\begin{tabular}{|p{7.5cm}|p{2.5cm}||p{2.5cm}|}
  \hline
   \hspace{2.6cm}Free parameters  & \hspace{1.1cm} $\omega$  & \hspace{1cm} $W$ \\[3mm]
   \hline
    $q = 0.5, \lambda = 0.5, c = 0.01, R_0 = 0.001,  f_{\epsilon} = g_{\epsilon} = 1.2 $ &\hspace{0.5cm} -1,+1,-1 & \hspace{1cm} -1 \\[3mm]
   \hline
    $ q = 0.5, \lambda = 0.5, c = 0.01, R_0 = 0.001,  f_{\epsilon} = g_{\epsilon} = 1.4 $ & \hspace{0.7cm} -1,+1 & \hspace{1.2cm}  0 \\[3mm]
  \hline
   $ q = 0.5, \lambda = 0.5, c = 0.01, R_0 = 0.001, f_{\epsilon} = g_{\epsilon} = 1.6$ & \hspace{1cm}  -1 & \hspace{1cm}  -1 \\[3mm]
  \hline
  $ q = 0.5, \lambda = 0.5, c = 0.01, R_0 = -0.001, f_{\epsilon} = g_{\epsilon} = 1.2 $ & \hspace{0.5cm} -1,+1,-1 & \hspace{1cm} -1  \\[3mm]
  \hline
  $q = 0.5, \lambda = 0.5, c = 0.01, R_0 = -0.001, f_{\epsilon} = g_{\epsilon} = 1.4 $ & \hspace{0.7cm} -1,+1 & \hspace{1.2cm} 0 \\[3mm]
  \hline
  $q = 0.5, \lambda = 0.5, c = 0.01, R_0 = -0.001, f_{\epsilon} = g_{\epsilon} = 1.6 $ & \hspace{1cm} -1 & \hspace{1.2cm} -1 \\[3mm]
  \hline
\end{tabular}
\caption{Summary of the results}\label{1}
\end{tabularx}
\end{table*}
\subsection{Photon sphere}
In our past papers \cite{sa31,sa32}, we demonstrated that the topological photon sphere can play a fundamental and reciprocal role in interpreting the behavior of a black hole. More precisely, since the existence of a photon sphere or photon rings is one of the essential requirements of ultra-compact gravitational structures, particularly black holes\cite{sa33}, the topological photon sphere can serve as a crucial test for determining the parameter ranges of black holes and, consequently, their overall behavior, and vice versa.
To achieve this, we first construct the regular H potential and the vector field $\phi$ \cite{sa34}.
\begin{equation}\label{1000}
\begin{split}
H(r, \theta)=\sqrt{\frac{-g_{tt}}{g_{\varphi\varphi}}},
\end{split}
\end{equation}
\begin{equation}\label{2000}
\begin{split}
&\phi^r=\frac{\partial_rH}{\sqrt{g_{rr}}},\\
&\phi^\theta=\frac{\partial_\theta H}{\sqrt{g_{\theta\theta}}},
\end{split}
\end{equation}
So with respect to Eq. (\ref{sol}), we will have,
\begin{equation}\label{5000}
H =\frac{\sqrt{36-\frac{36 m}{r}-\frac{3 r^{2} R_{0}}{g_{\varepsilon}^{2}}+\frac{36 \left(\frac{q^{2}}{r^{2}}-\frac{q^{4} \lambda}{20 r^{6}}\right) f_{\varepsilon}^{2}}{1+c}}\, g_{\varepsilon}}{6 \sin \! \left(\theta \right) r f_{\varepsilon}},
\end{equation}
\begin{equation}\label{6000}
\phi^{r}=\frac{3 g_{\varepsilon}^{2} \csc \! \left(\theta \right) \left(\frac{2 \left(-1-c \right) r^{6}}{3}+m \left(1+c \right) r^{5}-\frac{4 q^{2} r^{4} f_{\varepsilon}^{2}}{3}+\frac{2 q^{4} \lambda  f_{\varepsilon}^{2}}{15}\right)}{2 f_{\varepsilon} r^{8} \left(1+c \right)},
\end{equation}
\begin{equation}\label{7000}
\phi^{\theta}=-\frac{\sqrt{36-\frac{36 m}{r}-\frac{3 r^{2} R_{0}}{g_{\varepsilon}^{2}}+\frac{36 \left(\frac{q^{2}}{r^{2}}-\frac{q^{4} \lambda}{20 r^{6}}\right) f_{\varepsilon}^{2}}{1+c}}\, g_{\varepsilon}^{2} \cos \! \left(\theta \right)}{6 \sin \! \left(\theta \right)^{2} r^{2} f_{\varepsilon}}.
\end{equation}
Although our primary objective in this work is to examine the impact of gravitational corrections on the parameters of the Euler-Heisenberg model within the context of Rainbow Gravity, such as parameters $f_{\varepsilon}$, $g_{\varepsilon}$ , $c=f_{R_{0}}$ and $\lambda$, it appears that the parameter $R_{0}$ plays a more influential role. We will clearly observe that this parameter can affect the model's behavior and essentially simulate the model and its associated photon sphere as either dS or AdS. Therefore, in this section, we will examine this parameter in both its negative and positive forms. In each case, we will separately analyze the impact of the applied corrections on the black hole action.

\subsubsection{Case I: $R_{0}<0$}
\begin{figure*}[tbh!]
      	\centering{
       \includegraphics[scale=0.25]{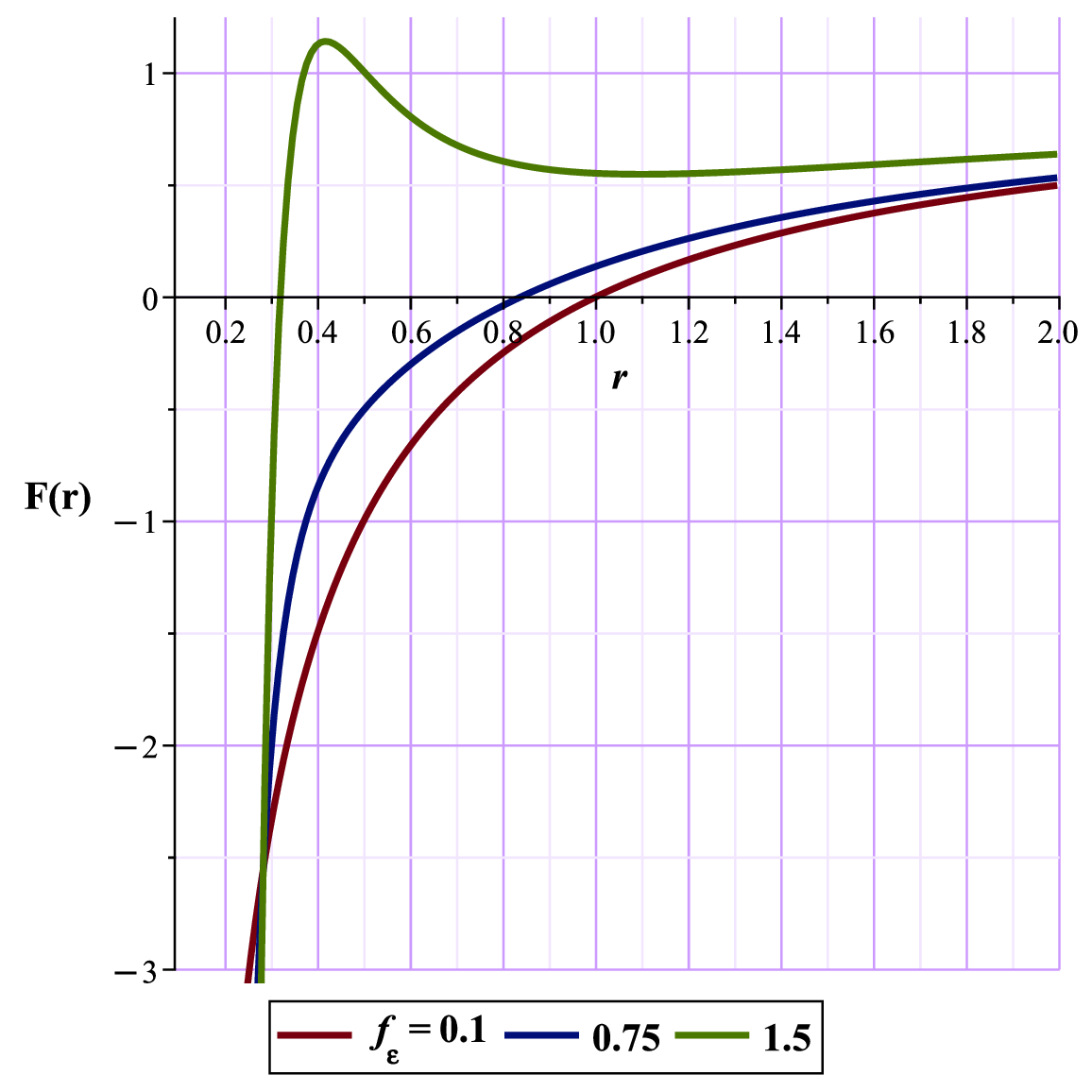} \hspace{2mm}
      }
       	\caption{Metric function with different $f_{\varepsilon}$  for the black hole in F(R)-Euler-Heisenberg gravity’s Rainbow}
      	\label{f5}
      \end{figure*}

In this scenario, according to the assumed values $m=1, R_{0}=-0.001, q=0.5, c=0.01, g_{\varepsilon}=1.1$ and $\lambda=0.5$ the model behaves entirely in the form of AdS, and only the event horizon appears, as shown in fig (\ref{f5}).
Although we have examined different values for the existing parameters, but since they exhibit similar behavior, we will present one case for consistency with the values chosen in the previous sections.
\begin{figure*}[tbh!]
      	\centering{
       \includegraphics[scale=0.25]{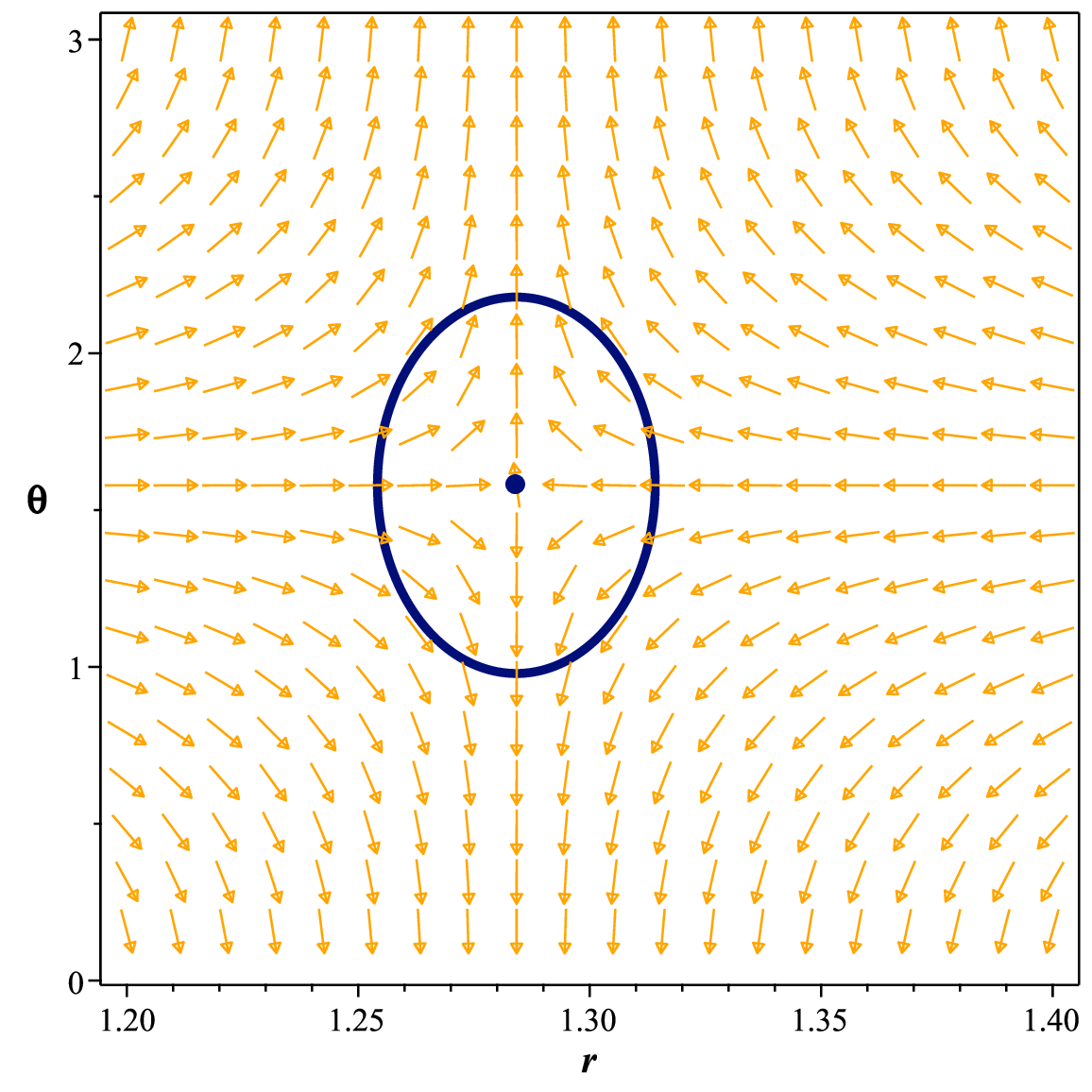}
      	\includegraphics[scale=0.25]{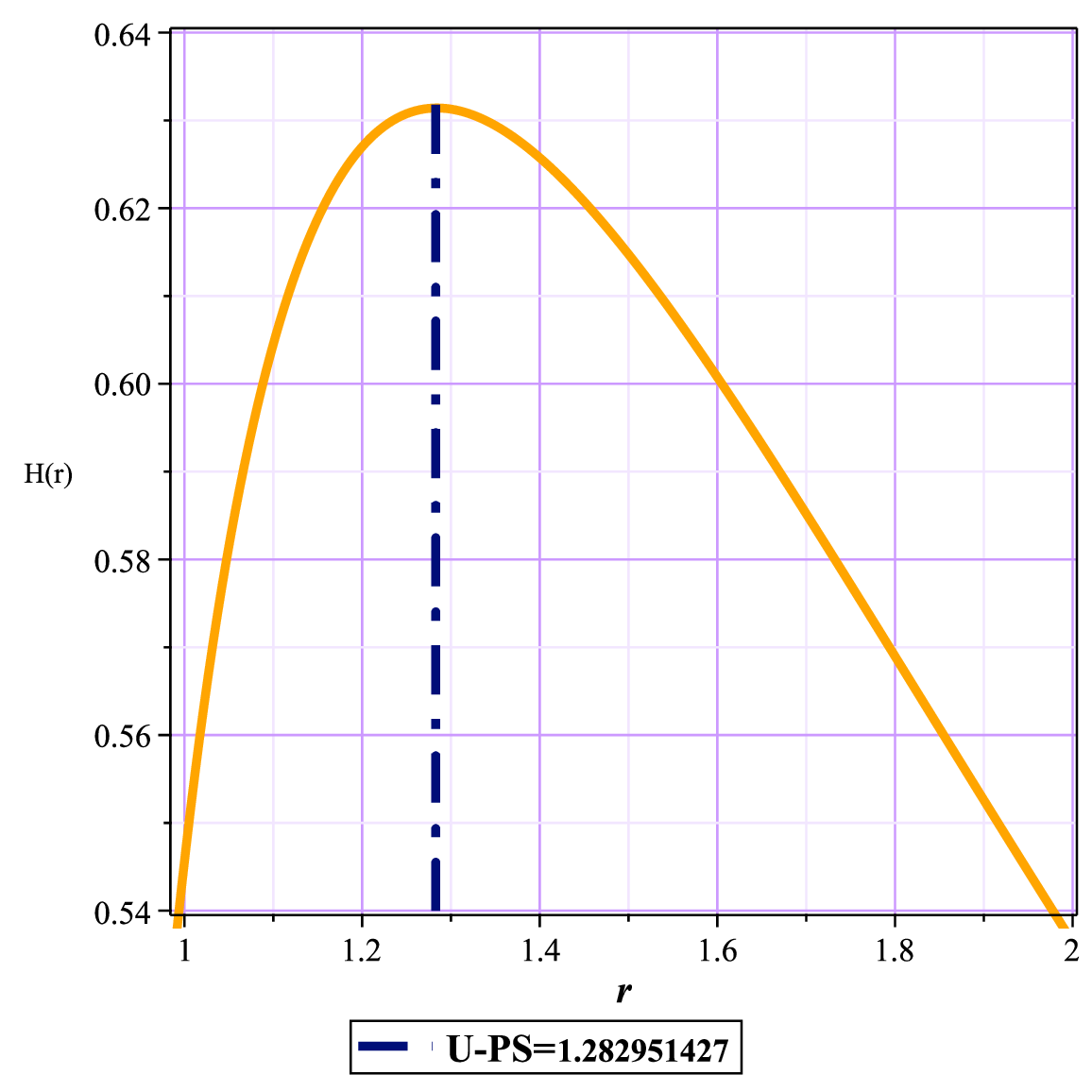}
      }
       	\caption{The normal vector field $n$ in the $(r-\theta)$ plane. The photon spheres are located at $ (r,\theta)=(1.284148559,1.57)$  with respect to $(\lambda=0.5,q=0.5,m=1,c=0.01,R_{0}=-0.001,g_{\varepsilon}=1.1),f_{\varepsilon}=0.75$, Right fig is the topological potential H(r) for black hole in F(R)-Euler-Heisenberg gravity’s Rainbow}
      	\label{f6}
      \end{figure*}
According to the method of determining topological charges \cite{sa31,sa32,sa35} , the  Figure (\ref{f6}) corresponds to a region where the system must possess an unstable photon sphere(U-PS) and a total topological charge of -1. This is confirmed by the potential function H(r) exhibiting a local maximum. Consequently, the structure will take the form of a normal black hole.Our investigations for various parameter values indicate that, in its AdS form and within the permissible parameter ranges, this black hole, unlike many previously studied black holes \cite{sa31,sa32,sa32.1,sa32.2,sa32.3}, does not exhibit the naked singularity region and consistently maintains the conditions of being a black hole for different parameter values.
\subsubsection{Case II: $R_{0}>0$}
\begin{figure*}[tbh!]
      	\centering{
       \includegraphics[scale=0.3]{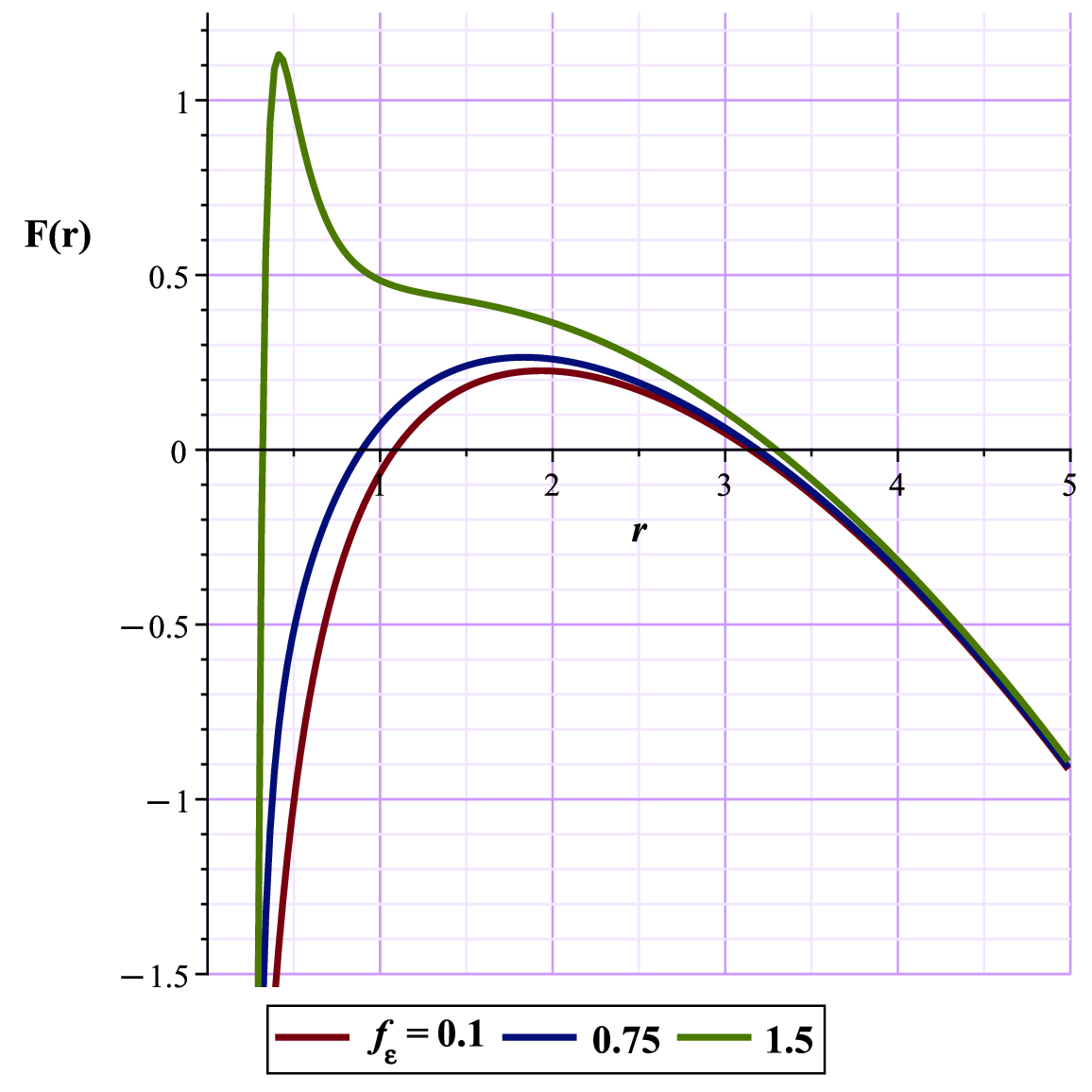} \hspace{2mm}
      }
       	\caption{Metric function with different  $f_{\varepsilon}$ for the black hole in F(R)-Euler-Heisenberg gravity’s Rainbow  }
      	\label{f8}
      \end{figure*}

By selecting positive values for $R_{0}$, the model changes its behavior, exhibiting a dS form. In this scenario, typically two or three horizons can appear: the Cauchy horizon, the event horizon, and the cosmological horizon. Based on our chosen parameters $m=1, R_{0}=1, q=0.5, c=0.01, g_{\varepsilon}=1.1$ and $\lambda=0.5$, it appears that only the event horizon and cosmological horizon are observed, as shown in fig (\ref{f8}).
Our previous studies on dS models \cite{sa31,sa32,sa32.1,sa32.2,sa32.3} indicate that due to the presence of the cosmological horizon, the model cannot exhibit any potential extremum beyond the cosmological horizon. In other words, it seems that no photon sphere model, whether stable or unstable, can emerge outside the cosmological horizon. Therefore, these photon spheres typically appear in the region between the event horizon and the cosmological horizon.
\begin{figure*}[tbh!]
      	\centering{
       \includegraphics[scale=0.3]{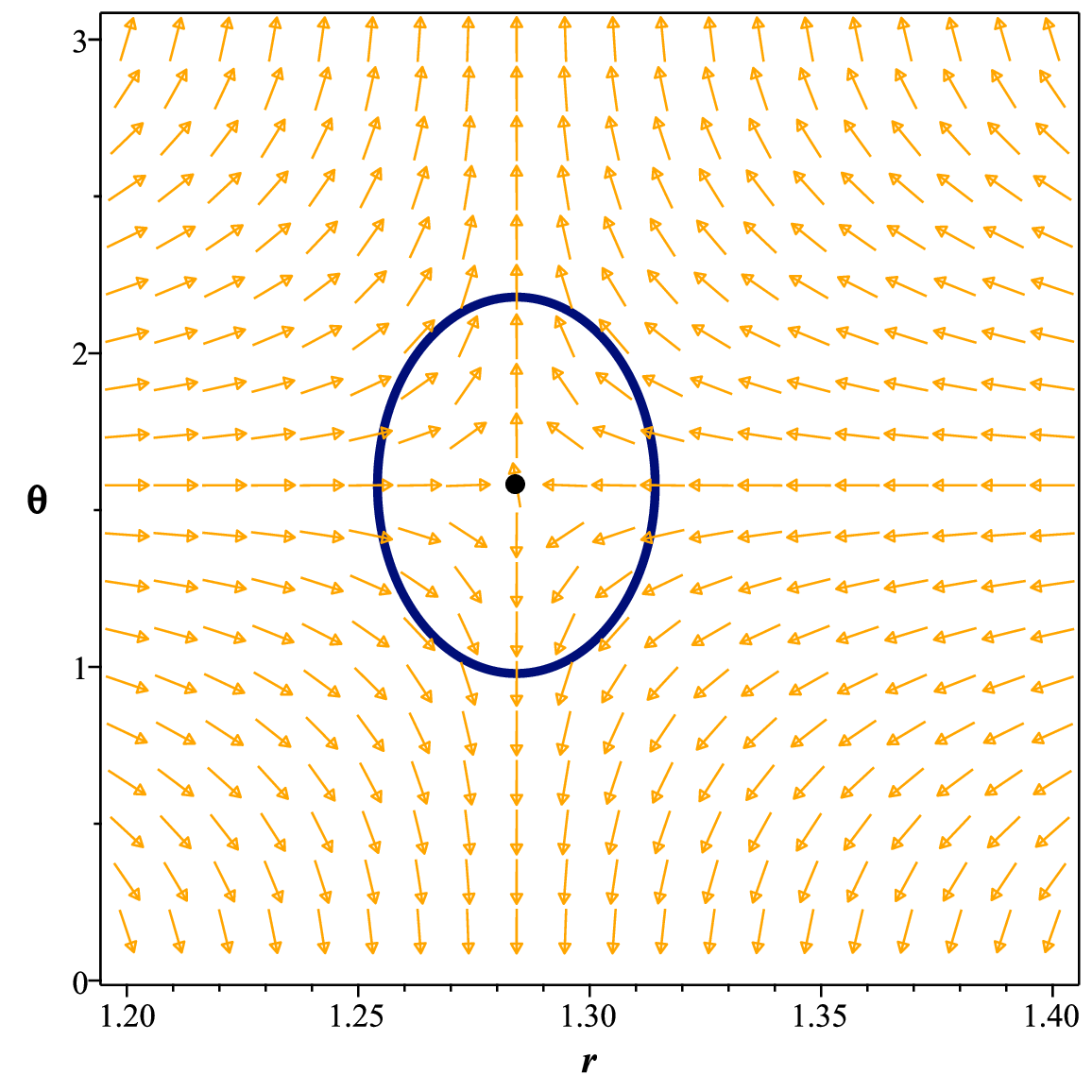} \hspace{2mm}
      	\includegraphics[scale=0.3]{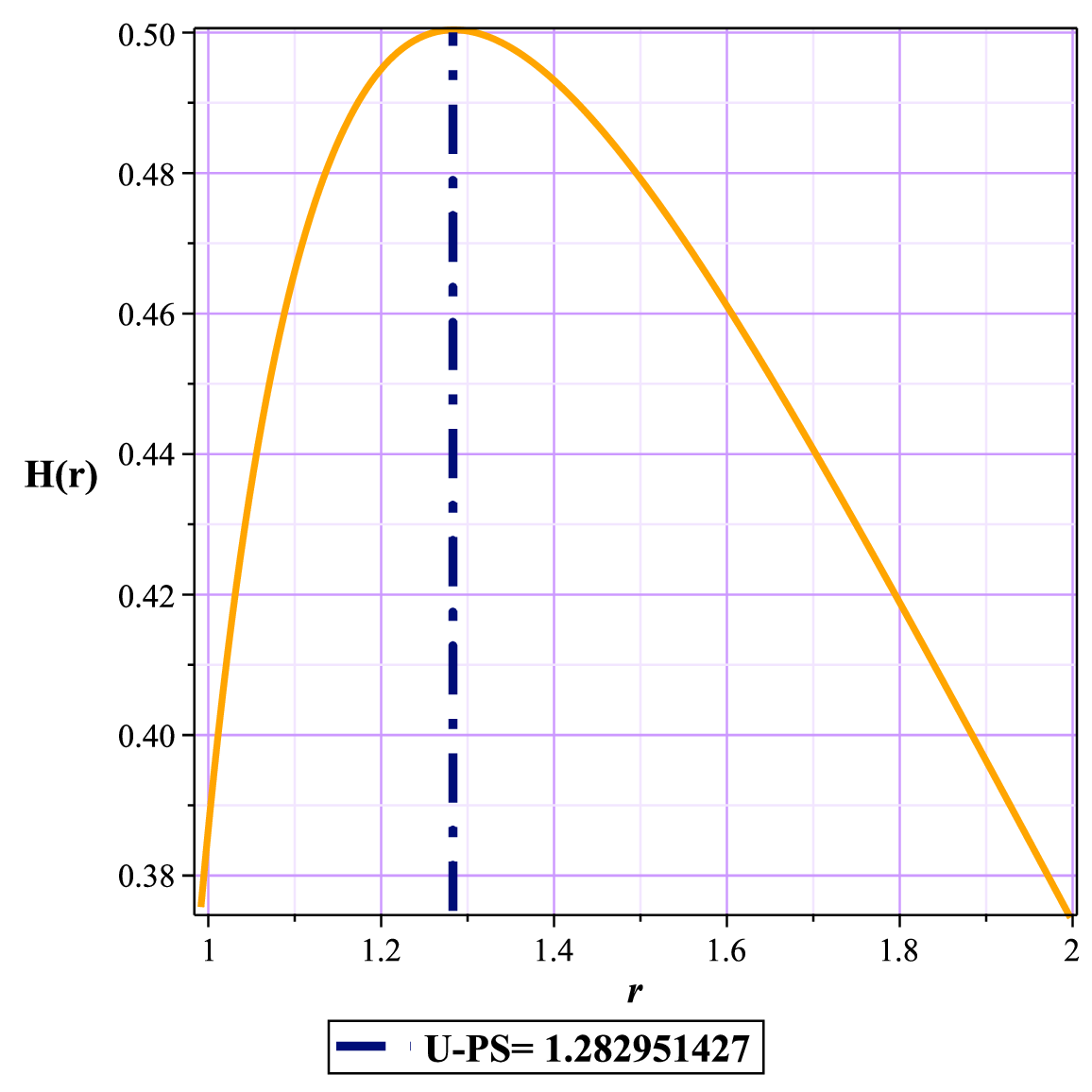} \hspace{2mm}
      }
       	\caption{The normal vector field $n$ in the $(r-\theta)$ plane. The photon spheres are located at $ (r,\theta)=(1.282951427,1.57)$ respect to with respect to $(\lambda=0.5,q=0.5,m=1,c=0.01,R_{0}=1,g_{\varepsilon}=1.1,f_{\varepsilon}=0.75)$, Right fig is the topological potential H(r) for Euler-Heisenberg black hole model}
      	\label{f9}
      \end{figure*}
In this context, the model, evaluated for multiple parameter values, consistently exhibits dS behavior and is unable to form a photon sphere outside the larger horizon or cosmological horizon for any parameter values. Consequently, as shown in fig (\ref{f9}), an unstable photon sphere with charge -1 always appears between the two horizons, corresponding to an energy maximum.
It is important to note that in this case, we selected a value for $R_{0}$ that provided the best representation. For instance, with $R_{0} \simeq 0.001$, the cosmological horizon would appear at $r=190 $, making its depiction in the figure somewhat challenging.
\section{Conclusion}
\label{conc}
The study of black hole thermodynamics delves into the intriguing connections between thermodynamic properties and the topological features of black holes. This field often involves examining critical points in black hole phase diagrams and assigning topological charges to these points. A notable approach is Duan's topological current $\phi$-mapping theory, which introduces the concept of topological charges to these critical points. This method offers a new perspective on the thermodynamic behavior of black holes and can be applied to other types of black holes, revealing more complex topological information. We have successfully derived the field equations for $F(R)$-Euler-Heisenberg theory, establishing a framework to explore the interaction between modified gravity and non-linear electromagnetic effects. Additionally, we have obtained an analytical solution for a static, spherically symmetric, energy-dependent black hole with constant scalar curvature. Also, our analysis of black holes in F(R)-Euler-Heisenberg gravity’s Rainbow provides significant insights into their topological properties. By examining the normalized field lines and various free parameters, we identified distinct winding and total topological numbers. Our findings suggest that the parameters $( R_0 )$ and $( f_{\epsilon} = g_{\epsilon} )$ influence the topological charges. The stability of these black holes was further assessed through winding numbers, highlighting the intricate relationship between these parameters and the black holes' topological characteristics. These results are comprehensively summarized in Table I. In the study of the photon sphere for this model, as emphasized above, the sign of the parameter$( R_0 )$ can significantly influence the model's behavior, transforming it into either a dS or AdS form. An interesting aspect of this model is that, in its AdS form, unlike many models where selecting different parameter values could behaviorally divide the space into black hole regions and naked singularities, this model does not exhibit a naked singularity region. It appears that, for various parameter values within the permissible range, the system can maintain its black hole behavior by displaying an unstable photon sphere. Ultimately, in the dS form, the behavior of the model's photon sphere remains general, like other dS models studied, and does not show any particular differences.
\label{conc}

\end{document}